\documentclass[letterpaper,11pt]{article}
\usepackage{jheppub}
\pdfoutput=1
\usepackage{slashed}
\usepackage{graphicx}
\usepackage{subfigure}
\usepackage{soul}

\usepackage{amsmath}
\usepackage{multirow}
\usepackage{tablefootnote}
\usepackage{hyperref}
\usepackage{epstopdf}
\usepackage{comment}
\usepackage{ulem}
\usepackage[utf8]{inputenc}

\usepackage[dvipsnames]{xcolor}

\newcommand{\eqna}{eqnarray}
\newcommand{\bea}{\begin{\eqna}}
\newcommand{\eea}{\end{\eqna}}
\newcommand{\subeq}{subequations}
\newcommand{\beqs}{\begin{\subeq}}
\newcommand{\eeqs}{\end{\subeq}}

\def\figureautorefname~#1\null{Fig.\,#1\null}
\def\tableautorefname~#1\null{Table\,#1\null}

\def\equationautorefname~#1\null{Eq.\,(#1)\null}

\def\m1{M_1}
\def\m2{M_2}
\def\m3{M_3}

\def\ch10{\tilde \chi^0_1}

\def\gev{\,{\rm GeV}}

\def\to{\rightarrow}

\newcommand{\lsim}{\mathrel{\mathop{\kern 0pt \rlap
  {\raise.2ex\hbox{$<$}}}
  \lower.9ex\hbox{\kern-.190em $\sim$}}}
\newcommand{\gsim}{\mathrel{\mathop{\kern 0pt \rlap
  {\raise.2ex\hbox{$>$}}}
  \lower.9ex\hbox{\kern-.190em $\sim$}}}

\definecolor{pink}{RGB}{255,105,180}

\def\cosba{\cos(\beta-\alpha)}
\def\cba{\cos(\beta-\alpha)}

\newcommand{\stu}{S,\ T\ \text{and}~ U}

\newcommand{\eehz}{e^+e^- \to hZ}

\newcommand{\eeww}{e^+e^- \to WW}
\newcommand{\tanb}{\tan \beta}
\newcommand{\lambvs}{\sqrt{\lambda v^2}}

\allowdisplaybreaks[4]

\title{Type-I 2HDM under the Higgs and Electroweak Precision Measurements }

\author[*]{Ning Chen,}
\author[\dagger]{Tao Han,}
\author[\diamond]{Shuailong Li,}
\author[\diamond]{Shufang Su,}
\author[\circ]{Wei Su,}
\author[\#]{Yongcheng Wu}

\affiliation[*]{School of Physics, Nankai University, Tianjin 300071, China}
\affiliation[\dagger]{Department of Physics and Astronomy, University of Pittsburgh,  Pittsburgh, PA 15260, USA}
\affiliation[\diamond]{Department of Physics, University of Arizona, Tucson, Arizona  85721, USA}
\affiliation[\circ]{ARC Centre of Excellence for Particle Physics at the Terascale, Department of Physics,University of Adelaide, South Australia 5005, Australia}
\affiliation[\#]{Ottawa-Carleton Institute for Physics, Carleton University, 1125 Colonel By Drive, Ottawa, Ontario K1S 5B6, Canada}

\preprint{PITT-PACC-1901}

\emailAdd{chenning$\_$symmetry@nankai.edu.cn, than@pitt.edu, shuailongli@email.arizona.edu, shufang@email.arizona.edu, wei.su@adelaide.edu.au, ycwu@physics.carleton.ca}

\abstract
 {We explore the extent to which future precision measurements of the Standard Model (SM) observables at the proposed $Z$-factories and Higgs factories may have impacts on new physics beyond the Standard Model, as illustrated by studying the Type-I Two-Higgs-doublet model (Type-I 2HDM). 
 We include the contributions from the heavy Higgs bosons at the tree-level and at the one-loop level in a full model-parameter space. While only small $\tanb$ region is strongly constrained at tree level,  the large $\tanb$ region gets constrained at loop level due to $\tan\beta$ enhanced tri-Higgs couplings.
 We perform a multiple variable $\chi^2$ fit with non-alignment and non-degenerate masses. We find  that the allowed parameter ranges could be tightly constrained by the future Higgs precision measurements, especially for small and large values of $\tan\beta$.  Indirect limits on the masses of heavy Higgs bosons can be obtained, which can be complementary to the direct searches of the heavy Higgs bosons at hadron colliders. We also find that the expected accuracies at the $Z$-pole and at a  Higgs factory are quite complementary in constraining mass splittings of heavy Higgs bosons.  The typical results are $|\cos(\beta-\alpha)| < 0.05,  |\Delta m_\Phi | < 200\ {\rm GeV}$, and $\tan\beta \gtrsim 0.3$. The reaches from CEPC, FCC-ee and ILC are also compared, for both Higgs and $Z$-pole precision measurements. Comparing to the Type-II 2HDM, the 95\% C.L. allowed range of $\cos(\beta-\alpha)$ is larger, especially for large values of $\tan\beta$.
  }
\keywords{Electroweak precision measurements, Higgs bosons,
Beyond the Standard Model, 2HDM.}

\begin{document}
\maketitle
\flushbottom
\clearpage


\section{Introduction}
\label{sec:intro}

The discovery of the Higgs boson at the CERN Large Hadron Collider \cite{Aad:2012tfa,Chatrchyan:2012xdj} has profound implications in our understanding of physics at short distances. It not only verifies the mechanism for the spontaneous electroweak symmetry breaking (EWSB), but also establishes a self-consistent theory, the ``Standard Model (SM)", that could be valid to an exponentially high scale, perhaps to the Planck Scale. 
Indeed, all the current measurements at the electroweak (EW) scale of a few hundred GeV seem to indicate the observed Higgs boson to be a SM-like elementary scalar. When high energy physics advances to the next level, it is thus a natural and pressing question to ask whether in Nature there are other Higgs bosons, associated with a new physics scale as predicted in many extended theories beyond the Standard Model (BSM). As such, searching for new Higgs bosons at the current and future facilities should be of high priority.

One of the well-motivated extensions is the 
Two-Higgs-doublet model (2HDM)~\cite{Branco:2011iw}. 
After the EWSB with the EW gauge bosons absorbing three Goldstone bosons, there are five massive spin-zero states left in the spectrum ($h, H,A,H^\pm$), among which $h$ is assumed to be the SM-like Higgs boson.\footnote{The case with the heavy CP-even Higgs $H$ being the SM-like Higgs is still consistent with the current experimental searches, although the viable parameter space has been tightly constrained when both direct and indirect search limits are combined.}
Extensive searches for the additional Higgs bosons have been actively carried out, especially at the LHC \cite{Aaboud:2017sjh,CMS-PAS-HIG-17-020,Aaboud:2017gsl,Aaboud:2017rel,Sirunyan:2018qlb,Aaboud:2017yyg,Aaboud:2017cxo,Aaboud:2018eoy,Khachatryan:2016are,Aaboud:2018ftw,Sirunyan:2018iwt,ATLAS-CONF-2016-089,Aaboud:2018gjj,CMS-PAS-HIG-16-031,Sirunyan:2019wph,Sirunyan:2019tkw,Aaboud:2019sgt,Sirunyan:2018taj, CMS:2019hvr,Aad:2019zwb}. 
In the absence of signal observation at the LHC experiments, this would imply either the new Higgs bosons are much heavier and essentially decoupled from the SM, or their interactions with the SM particles are highly suppressed and the couplings of the SM-like Higgs accidentally aligned with the SM predictions~\cite{Carena:2013ooa,Dev:2014yca}.  
In either situation, it would be challenging to directly produce those states in the current and near-future experiments.

Alternatively, precision measurements of SM observables and the Higgs properties could lead to relevant insights into new physics. The recent proposals of construction of a Higgs factory, including the Circular Electron Positron Collider (CEPC) in China~\cite{CEPC-SPPCStudyGroup:2015csa,CEPCStudyGroup:2018ghi}, the electron-positron stage of the Future Circular Collider (FCC-ee) at CERN (previously known as TLEP~\cite{Gomez-Ceballos:2013zzn,fccpara,fccplan}), and the International Linear Collider (ILC) in Japan~\cite{Baer:2013cma,Bambade:2019fyw,Fujii:2019zll}, could shed light on new physics in the pursuit of precision Higgs measurements. 
With about $10^6$ Higgs bosons expected at the Higgs factory, one would be able to reach sub-percentage precision determination of the Higgs properties, and thus to be sensitive to new physics associated with the Higgs boson. 
As an integrated part of the circular collider program, one would like to return to the $Z$-pole. 
With about $10^{10} - 10^{12}\ Z$ bosons, the achievable precisions on the SM observables could be improved by a factor of $20-200$ over the Large Electron Positron (LEP) Collider results~\cite{ALEPH:2005ab}. 
Such an unprecedented precision would hopefully lead to hints of new physics associated with the EW sector.

There is a plethora of articles in the literature to study the effects of the heavy Higgs states on the SM observables~\cite{Branco:2011iw}. In particular, a few current authors performed a study focusing on the Type-II 2HDM~\cite{Chen:2018shg}. We found that the expected accuracies at the $Z$-pole and at a Higgs factory are quite constraining to mass splittings of heavy Higgs bosons. The reach in the heavy Higgs masses and couplings can be complementary to the direct searches of the heavy Higgs bosons at hadron colliders. In this paper, we extend the previous study by examining  the Type-I 2HDM. There are interesting and qualitative differences in those two models. One of the most distinctive features comes from the coupling pattern of the Higgs bosons to the SM fermions. Relevant to our studies are the Yukawa couplings of the SM-like Higgs boson $h$. The deviations from the SM predictions scale with a factor $\cot\beta$ in Type-I, while the scaling factor for down-type fermions goes like $\tan\beta$ in Type-II.
In our analyses, we include the tree-level corrections to the SM-like Higgs couplings and one-loop level contributions from heavy Higgs bosons. We perform a $\chi^2$ fit in the full model-parameter space. In particular, we study the extent to which the parametric deviations from the alignment and degenerate masses can be probed by the precision measurements.  We will comment on the results whenever there is a difference between Type-I and Type-II.

The rest of the paper is organized as follows. 
An overview of the Higgs and electroweak precision observables at future $e^+ e^-$ colliders is given in~\autoref{sec:input}, which will serve as inputs for our analyses of the Type-I 2HDM.
We present the Type-I 2HDM and the one-loop corrections in~\autoref{sec:model}. 
In~\autoref{sec:theory}, we impose the set of theoretical constraints to the Higgs boson masses and self-couplings.
In~\autoref{sec:LHC}, the constraints from the direct LHC searches for heavy Higgs bosons are presented for the Type-I 2HDM, under the current LHC runs and the future projected HL-LHC sensitivities. 
\autoref{sec:results} shows our main results from the $\chi^2$ fit, for the cases of mass degeneracy and non-degeneracy of heavy Higgs bosons. We summarize our results and draw conclusions in~\autoref{sec:conclu}.
Some useful analytic formulae and approximate treatments are given in~\autoref{sec:app}.


\section{Higgs Observables at Future Lepton Colliders}
\label{sec:input}

The SM has been tested to a high precision from the measurements at the $Z$-pole from LEP-I \cite{ALEPH:2005ab}, at  the Tevatron \cite{Baak:2014ora} and the LHC \cite{Haller:2018nnx}. 
It has been demonstrated that the EW and Higgs precision measurements  can impose strong constraints on new physics models~\cite{Gori:2015nqa,Su:2016ghg}. In this section, we closely follow the approach adopted in Ref.~\cite{Chen:2018shg}, and list the projected precision achievable by several proposed future $e^+e^-$ machines on $Z$-pole and Higgs measurements. 


\begin{table}[!tb]
    \centering
    \resizebox{0.6 \textwidth}{!}{
    \begin{tabular}{c|l|l|l}
    \hline\hline
        Observables & FCC-ee & CEPC & ILC \\
        \hline\hline
        $\delta m_h$ [GeV] & $1.0\times10^{-2}$ & $5.9 \times10^{-3}$& $1.5\times10^{-2}$\\
        $\delta\alpha_{\rm had}$ & $3.8\times10^{-5}$ * & $4.7\times10^{-5}$& $3.8\times10^{-5}$ *\\
        $\delta m_Z$ [GeV] & $1.0\times10^{-4}$ & $5.0\times10^{-4}$  & 
        $2.1\times10^{-3}$  \\
        $\delta m_t$ [GeV] & $2.0\times10^{-2}$  & $6.0\times10^{-1}$  & $1.7\times10^{-2}$ \\
        %
        $\delta m_W$ [GeV] & $7.0\times10^{-4}$ & $1.0\times10^{-3}$  & $2.5\times10^{-3}$\\
        $\delta \Gamma_W$ [GeV] & $1.5\times10^{-3}$  & $2.8\times10^{-3}$ & $5.0\times10^{-3}$\\
        $\delta \Gamma_Z$ [GeV] & $1.0\times10^{-4}$  & $5.0\times10^{-4}$ & 
        $7.0\times10^{-4}$\\
        $\delta A_b^{\rm FB}$ & $3.0\times10^{-4}$ & $1.0\times10^{-4}$ & $1.6\times10^{-3}$ *\\
        $\delta A_c^{\rm FB}$ & $5.9\times10^{-4}$ & $2.2\times10^{-4}$ & $3.5\times10^{-3}$ * \\
        $\delta A_\ell^{\rm FB}$ & $9.0\times10^{-6}$&  $5.0\times10^{-5}$ & $1.0\times10^{-3}$ *\\
        %
        %
        $\delta R_b$ & $6.0\times10^{-5}$ & $4.3\times10^{-5}$   & $1.5\times10^{-4}$  \\
        $\delta R_c$ & $1.7\times10^{-4}$ & $1.7\times10^{-4}$ & $5.2\times10^{-4}$ \\
        $\delta R_\ell$ & $1.0\times10^{-3}$ & $2.1\times10^{-3}$ & $4.0\times10^{-3}$ \\
        $\delta \sigma_{\rm had}$ [nb] & $4.0\times10^{-3}$ & $5.0\times10^{-3}$ & $3.7\times10^{-2}$ * \\
        %
        %
        \hline\hline
    \end{tabular}
    }
    \caption{The observables and corresponding precision used in $S/T/U$ fitting for each future collider. Most of the values~\cite{deBlas:2019rxi} come from the corresponding CDRs of FCC-ee~\cite{Abada:2019lih,Abada:2019zxq}, CEPC~\cite{CEPCStudyGroup:2018ghi}, and ILC~\cite{Fujii:2019zll},  except  for the values  with *, which comes from earlier studies~\cite{Chen:2018shg,Fan:2014vta,Baak:2013fwa}.  For ILC we choose its Giga-Z scenario.}
    \label{tab:ew_obs}
\end{table}

These measurements are expected to be significantly improved by a new run at the $Z$-pole at future lepton colliders with a much larger data sample~\cite{CEPCPhysics-DetectorStudyGroup:2019wir,CEPC-SPPCStudyGroup:2015csa,Gomez-Ceballos:2013zzn,Asner:2013psa,fccplan,fccpara}.   
The expected precision on the measurements of $\Delta \alpha_{\rm had}^{(5)} (M_Z^2)$, $m_Z$, $m_t$, $m_h$, $m_W$,  $\Gamma_Z$ et al. are summarized in~\autoref{tab:ew_obs}\footnote{ Entries of our~\autoref{tab:ew_obs} are mostly the same as those in Table 27 of Ref.~\cite{deBlas:2019rxi}, except  there is one typo of $\delta A_\ell^{\rm FB}$ which was confirmed with the authors. Other small differences appear when we use the values from CDRs, while Ref.~\cite{deBlas:2019rxi} updated a few based on some private discussions.  }. Here  we take the Giga-$Z$ plan for the ILC $Z$-pole running.

In~\autoref{tab:STU}, we show the current~\cite{Haller:2018nnx} as well as the predicted precisions on the oblique parameters at future lepton colliders, together with the correlation error matrix. For the predicted precisions for future machines, {\tt Gfitter} package~\cite{Baak:2014ora} is used with the precisions of electroweak measurements in~\autoref{tab:ew_obs}.  
In our analyses as detailed in a later section, the $\stu$ contours at  $95\%$ Confidence Level (C.L.) are adopted to constrain the 2HDM parameter spaces,  using the $\chi^2$  profile-likelihood  fit with error-correlation matrix.  Compared to the previous study in Ref.~\cite{Chen:2018shg}, the updated $\stu$ in Table \ref{tab:STU} lead to stronger constraints because of the strong correlations with large off-diagonal elements in the correlation matrices. 

\begin{table}[tb]
\centering
\resizebox{\textwidth}{!}{
  \begin{tabular}{|l|c|r|r|r|c|r|r|r|c|r|r|r|c|r|r|r|c|r|r|r|}
   \hline
    & \multicolumn{4}{c|}{Current}& \multicolumn{4}{c|}{CEPC}& \multicolumn{4}{c|}{FCC-ee }&\multicolumn{4}{c|}{ILC} \\
   \hline
   \multirow{2}{*}{}
   &\multirow{2}{*}{$\sigma$} &\multicolumn{3}{c|}{correlation}
   &{$\sigma$} &\multicolumn{3}{c|}{correlation}
   &{$\sigma$} &\multicolumn{3}{c|}{correlation}
   &{$\sigma$} &\multicolumn{3}{c|}{correlation} \\
   \cline{3-5}\cline{7-9}\cline{11-13}\cline{15-17}
   &&$S$&$T$&$U$&($10^{-2}$)&$S$&$T$&$U$&($10^{-2}$)&$S$&$T$&$U$&($10^{-2}$)&$S$&$T$&$U$\\
   \hline
   $S$& $0.04 \pm 0.11$& $1$ & $0.92$ & $-0.68$ & $1.82$  & $1$     & $0.9963$       & $-0.9745$ &   $0.370$    &  $1$     &   $0.9898$    &    $-0.8394$   &   $2.57$    &   $1$    &    $0.9947$   & $-0.9431$ \\
\hline
   $T$&$0.09\pm 0.14$& $-$ & $1$ & $-0.87$ & $2.56$  &  $-$   &  $1$      &  $-0.9844$   &   $0.514$    &   $-$    &    $1$   &    $-0.8636$   &    $3.59$   &   $-$    &   $1$    &   $-0.9569$\\
\hline
   $U$& $-0.02 \pm 0.11$& $-$ & $-$ & $1$ &$1.83$  &  $-$   &  $-$     &  $1$   &   $0.416$    &   $-$    &   $-$    &    $1$   &  $2.64$     &   $-$    &   $-$    & $1$ \\
   \hline
  \end{tabular}
  }
  \caption{Estimated $S$, $T$, and $U$ ranges and correlation matrices $\rho_{ij}$  from $Z$-pole precision measurements of the current results~\cite{Haller:2018nnx}, mostly from LEP-I~\cite{ALEPH:2005ab},  and at future lepton colliders at \autoref{tab:ew_obs}.  {\tt Gfitter} package~\cite{Baak:2014ora} is used in obtaining those constraints. }
\label{tab:STU}
\end{table}

A Higgs factory with $e^+e^-$ collisions at a  center-of-mass energy of $240-250$ GeV exploits the Higgsstrahlung process
\begin{equation}
\eehz .
\end{equation}
Owing to the clean experimental condition and well-constrained kinematics at the lepton colliders, both the inclusive cross section $\sigma(hZ)$ independent of the Higgs decays, and the exclusive channels of individual Higgs decays in terms of $\sigma(hZ)\times {\rm BR}$, can be measured to remarkable precisions.  The invisible decay width of the Higgs boson can also be sensitively probed. In addition, the  cross sections of vector boson fusion processes for the Higgs production ($WW,ZZ\to h$) grow with the center of mass energy logarithmically.  While their rates are still rather small at 240-250 GeV, at higher energies in particular for a linear collider,  such fusion processes become significantly more important and can provide crucial complementary information. For $\sqrt{s}>500$ GeV, $t \bar t h$ production can also be utilized as well.

We list the running scenarios of various machines in terms of their center of mass energies and the corresponding integrated luminosities, as well as the estimated precisions of relevant Higgs measurements that are used in our global analyses in Table~\ref{tab:mu_precision}.  
These expected results in the table serve as the input values for the later studies in this paper in constraining the theoretical parameters in the BSM Higgs sector.
Comparing to the values used in earlier study of Ref.~\cite{Chen:2018shg}, the main update is the $h \to \gamma \gamma$ precision at  FCC-ee, which is 9\% instead of 4\% because of different simulation methods \cite{Benedikt:2651299}. 
We only include the rate information for the Higgsstrahlung $Zh$ and  the $WW$ fusion process in our $\chi^2$ fit. Some other measurements, such as the angular distributions, the diboson process $\eeww$, can provide additional information in addition to the rate measurements alone~\cite{Beneke:2014sba,Craig:2015wwr,Durieux:2017rsg}.

Future high-energy lepton colliders will have the capacity to perform precision measurements for the SM parameters, as already presented for a 1-TeV ILC~\cite{Fujii:2019zll} and multiple TeV CLIC~\cite{deBlas:2018mhx,Roloff:2018dqu}. On the other hand, the most important aspect of those machines will be to reach a higher energy threshold and thus likely to directly explore new physics beyond the SM. In the context of 2HDM, the BSM Higgs sector would be more readily probed at those machines by direct searches via processes like $e^+e^- \to Z^*\to  b\bar bA/H,\ AH$ and $H^+H^-$, etc. Clearly, those studies would be interesting and important, but the analyses of the signals and backgrounds would be quite a different task from the current focus based on the Higgs and EW precision measurements on the SM parameters.

Non-oblique corrections to $Zf\bar{f}$ vertex could also be used to constrain the contributions from the non-SM Higgs sector.  In particular, $R_b$ and $A_{FB}^b$ will be measured with high precision at future lepton colliders~\cite{deBlas:2019rxi}.  The reach in the charged Higgs boson mass and $\tan\beta$ is comparable to the Higgs precision measurements~\cite{Haber:1999zh}, which are complementary to the oblique corrections that are more sensitive to the mass differences between the charged Higgs and the neutral ones.


\begin{table}[tb]
 \begin{center}
  \begin{tabular}{|l|r|r|r|r|r|r|r|r|r|r|}
   \hline
   collider& \multicolumn{1}{c|}{CEPC}& \multicolumn{3}{c|}{FCC-ee}&\multicolumn{5}{c|}{ILC} \\
   \hline
   $\sqrt{s}$     &  $\text{240\,GeV} $ &  $\text{240\,GeV}$	&\multicolumn{2}{c|}{$\text{365\,GeV}$}  &  \text{250\,GeV}  &
   \multicolumn{2}{c|}{\text{350\,GeV}}  & \multicolumn{2}{c|}{\text{500\,GeV}} \\
   $\int{\mathcal{L}}dt $     &  $\text{5.6 ab}^{-1} $ &  $\text{5 ab}^{-1}$	&
   \multicolumn{2}{c|}{$\text{1.5 ab}^{-1}$}      &  $\text{2 ab}^{-1} $  &
   \multicolumn{2}{c|}{$\text{200 fb}^{-1}$}  & \multicolumn{2}{c|}{$\text{4 ab}^{-1}$} \\
   \hline
    \hline
production& $Zh$  & $Zh$   & $Zh$   &$\nu\bar{\nu}h$     & $Zh$      & $Zh$     & $\nu\bar{\nu}h$     & $Zh$     & $\nu\bar{\nu}h$  \\
   \hline
  $\Delta \sigma / \sigma$ & 0.5\%  & 0.5\%& 0.9\% &$-$ & 0.71\% & 2.0\% & $-$ & 1.05 & $-$ \\ \hline \hline
   decay & \multicolumn{9}{c|}{$\Delta (\sigma \cdot BR) / (\sigma \cdot BR)$}  \\
  \hline
   $h \to b\bar{b}$              &  0.27\%               & 0.3\%       			  &  0.5\%			& 0.9\%		           &  0.46\%    & 1.7\%         & 2.0\%                  & 0.63\%    & 0.23\%         \\

   $h \to c\bar{c}$              & 3.3\%                   & 2.2\%                   &6.5\%			&10\%					& 2.9\%          & 12.3\%    & 21.2\%                   & 4.5\%     & 2.2\%                    \\

   $h \to gg$                    & 1.3\%                   & 1.9\%                  & 3.5\%           & 4.5\%    & 2.5\%           & 9.4\%    & 8.6\%                  & 3.8\%     & 1.5\%                  \\

   $h \to WW^*$                  & 1.0\%                   & 1.2\%                   & 2.6\%          & 3.0\%    & 1.6\%          & 6.3\%    & 6.4\%                  & 1.9\%    & 0.85\%               \\

   $h \to \tau^+\tau^-$         & 0.8\%                   & 0.9\%                   &1.8\%           &8.0\%          &1.1\%           &4.5\%          & 17.9\%                  & 1.5\%    & 2.5\%               \\

   $h \to ZZ^*$                  & 5.1\%                   & 4.4\%                   & 12\%        & 10\%     & 6.4\%        & 28.0\%     & 22.4\%                   & 8.8\%     & 3.0\%                \\

   $h \to \gamma\gamma$          & 6.8\%                   & 9.0\%                   & 18\%     & 22\%     & 12.0\%     & 43.6\%     &50.3\%                   & 12.0\%   &6.8\% \\

   $h \to \mu^+\mu^-$           & 17\%                   & 19\%                  & 40\%        & $-$     & 25.5\%        & 97.3\%     & 178.9\%                  & 30.0\%     & 25.0\%                    \\
   \hline
    $(\nu\bar\nu)h \to b\bar{b}$  & 2.8\%       &   3.1\%      & $-$  & $-$ &       3.7\% & $-$  & $-$  & $-$  & $-$   \\
    \hline
  \end{tabular}
  \caption{Estimated statistical precisions for Higgs measurements obtained at  the proposed CEPC program with 5.6 ab$^{-1}$ integrated luminosity~\cite{CEPCStudyGroup:2018ghi,CEPCPhysics-DetectorStudyGroup:2019wir}, FCC-ee program with 5 ab$^{-1}$ integrated luminosity~\cite{Abada:2019lih,Abada:2019zxq},  and  ILC with various center-of-mass energies~\cite{Bambade:2019fyw}. 
   }
\label{tab:mu_precision}
  \end{center}
\end{table}
%

\section{Type-I Two-Higgs-Doublet Model}
\label{sec:model}

A generic 2HDM consists of two ${\rm SU}(2)_L$ scalar doublets $\Phi_i\ (i=1,2)$ with a hyper-charge assignment $Y=+1/2$
\begin{equation}
\Phi_{i}=\begin{pmatrix}
  \phi_i^{+}    \\
  (v_i+\phi^{0}_i+iG_i)/\sqrt{2}
\end{pmatrix}\,.
\end{equation}
After the EWSB, each doublet obtains a vacuum expectation value (vev)  $v_i\ (i=1,2)$ with $v_1^2+v_2^2 = v^2 = (246\ {\rm GeV})^2$, and  $v_2/v_1=\tan\beta$.

The 2HDM Lagrangian for the Higgs sector is given by
\begin{equation}\label{equ:Lall}
\mathcal{L}=\sum_i |D_{\mu} \Phi_i|^2 - V(\Phi_1, \Phi_2) + \mathcal{L}_{\rm Yuk}\,,
\end{equation}
with the CP-conserving potential 
\begin{eqnarray}
\label{eq:L_2HDM}
 V(\Phi_1, \Phi_2) &=& m_{11}^2\Phi_1^\dag \Phi_1 + m_{22}^2\Phi_2^\dag \Phi_2 -m_{12}^2(\Phi_1^\dag \Phi_2+ h.c.) + \frac{\lambda_1}{2}(\Phi_1^\dag \Phi_1)^2 + \frac{\lambda_2}{2}(\Phi_2^\dag \Phi_2)^2  \notag \\
 & &+ \lambda_3(\Phi_1^\dag \Phi_1)(\Phi_2^\dag \Phi_2)+\lambda_4(\Phi_1^\dag \Phi_2)(\Phi_2^\dag \Phi_1)+\frac{\lambda_5}{2}   \Big[ (\Phi_1^\dag \Phi_2)^2 + h.c.\Big]\,,
\end{eqnarray}
and a soft $\mathbb{Z}_2$ symmetry breaking term $m_{12}^2$.

One of the four neutral components and two of the four charged components are eaten by the SM gauge bosons $Z$, $W^\pm$ after the EWSB, providing their masses. The remaining physical mass eigenstates are two CP-even neutral Higgs bosons $h$ and $H$, with $m_h<m_H$, one CP-odd neutral Higgs boson $A$, plus a pair of charged Higgs bosons  $H^\pm$. Instead of the eight parameters appearing in the Higgs potential $m_{11}^2, m_{22}^2, m_{12}^2, \lambda_{1,2,3,4,5}$, a more convenient set of the parameters is $v, \tan\beta, \alpha, m_h, m_H, m_A, m_{H^\pm}, m_{12}^2$, where $\alpha$ is the rotation angle diagonalizing the CP-even Higgs mass matrix. We choose $ m_h = 125$ GeV to be the SM-like Higgs boson.

The Type-I 2HDM is characterized by the choice of the Yukawa couplings to the SM fermions and they are of the form 
 \begin{equation}\label{eq:Lyuk}
  -\mathcal{L}_{\rm Yuk}=Y_{d}{\overline Q}_L\Phi_2d_R^{}+Y_{e}{\overline L}_L\Phi_2 e_R^{}+ Y_{u}{\overline Q}_Li\sigma_2\Phi^*_2u_R^{}+\text{h.c.}\,.
\end{equation}
After the EWSB, the effective Lagrangian for the light CP-even Higgs couplings to the SM particles can be parameterized as
 \begin{eqnarray}\label{eq:Leff}
&&\mathcal{L}= \kappa_Z \frac{m_Z^2}{v}Z_{\mu}Z^{\mu}h+\kappa_W \frac{2m_W^2}{v}W_{\mu}^+ W^{\mu-}h + \kappa_g \frac{\alpha_s}{12 \pi v} G^a_{\mu\nu}G^{a\mu\nu}h + \kappa_{\gamma} \frac{\alpha}{2\pi v} A_{\mu\nu}A^{\mu\nu} h \nonumber \\
&& + \kappa_{Z\gamma}\frac{\alpha}{\pi v}A_{\mu\nu}Z^{\mu\nu}h -\Big( \kappa_u \sum_{f=u,c,t} \frac{m_f}{v}f \bar f + \kappa_d \sum_{f=d,s,b} \frac{m_f}{v}f \bar f + \kappa_{e} \sum_{f=e,\mu,\tau} \frac{m_f}{v}f \bar f \Big)h\,,
\end{eqnarray}
where
\begin{equation}
\label{eq:alias}
\kappa_i = \frac{g_{hii}^{\rm BSM}}{g_{hii}^{\rm SM}}\,,
\end{equation}
with $i$ indicating the individual Higgs coupling. Their values at the tree level are
\begin{equation}
\label{eq:kap_tree}
\kappa_Z^{\rm tree}=\kappa_W^{\rm tree}=\sin(\beta-\alpha)\,,\quad \kappa_f^{\rm tree}=\frac{\cos\alpha}{\sin\beta}=\sin(\beta-\alpha)+\cosba \cot \beta\,.
\end{equation}
We adopt the sign convention $\beta \in (0, \pi/2)$, $\beta -\alpha \in [0, \pi]$, so that $\sin (\beta-\alpha) \geq 0$.  Note that comparing to the Type-II 2HDM, in which up-type Yukawa couplings are proportional to $\cot\beta$ and bottom-type and lepton Yukawa couplings are proportional to $\tan\beta$, all the tree-level Yukawa couplings in the Type-I 2HDM are   proportional to $\cot\beta$. 
Therefore, $\kappa_f^{\rm tree}$ are enhanced comparing to the SM values only at low $\tan\beta<1$ region.

At the leading order, the CP-even Higgs couplings to the SM gauge bosons are $g_{hVV} = \sin(\beta-\alpha)$, and $g_{HVV} = \cos(\beta-\alpha)$.
The current measurements of the Higgs boson properties from the LHC are consistent with the SM Higgs boson interpretation. There are two well-known limits in 2HDM that would lead to a SM-like Higgs sector. 
The first situation is the alignment limit~\cite{Carena:2013ooa, Bernon:2015qea} of $\cos(\beta-\alpha)=0$, in which the light CP-even Higgs boson couplings are identical to the SM ones, regardless of the other scalar masses, potentially leading to rich BSM physics. 
For $\sin(\beta-\alpha)=0$, the opposite situation occurs with the heavy $H$ being identified as the SM Higgs boson. While it is still a viable option for the heavy Higgs boson being the observed 125 GeV SM-like Higgs boson~\cite{Coleppa:2014cca, Bernon:2015wef}, the allowed parameter space is being squeezed with the stringent direct and indirect experimental constraints. 
Therefore, in our analyses below, we identify the light CP-even Higgs $h$ as the SM-like Higgs with $m_h$ fixed to be 125 GeV. The other well-known case is the ``decoupling limit'', in which the heavy mass scales are all large: $m_{A, H, H^\pm}\gg 2m_Z$~\cite{Haber:1994mt}, so that they decouple from the low energy spectrum.  For masses of heavy Higgs bosons much larger than $\lambda_i v^2$, $\cos(\beta-\alpha)\sim \mathcal{O}(m_Z^2 / m_A^2)$  under perturbativity and unitarity requirement. Therefore, the light CP-even Higgs boson $h$ is again SM-like. Although it is easier and natural to achieve the decoupling limit by taking all the other mass scales to be heavy, there would be little BSM observable effects given the nearly inaccessible heavy mass scales. We will thus mainly focus on the alignment limit.

While the couplings $hgg$, $h\gamma\gamma$ and $h{Z\gamma}$ are absent at the tree-level in both the SM and the 2HDM, they are generated at the loop-level. 
In the SM, $hgg$, $h\gamma\gamma$ and $h{Z\gamma}$ all receive contributions from fermions (mostly top quark) running in the loop, while $h\gamma\gamma$ and $h{Z\gamma}$ receive contributions from $W$-loop in addition~\cite{Henning:2014wua}.
In 2HDM, the corresponding $hff$ and $hWW$ couplings that enter  the loop corrections need to be modified to the corresponding 2HDM values. Expressions for the dependence of $\kappa_g$, $\kappa_\gamma$ and $\kappa_{Z\gamma}$ on $\kappa_V$ and $\kappa_f$ can be found in Ref.~\cite{Heinemeyer:2013tqa}. 
There are, in addition, loop corrections to $\kappa_g$, $\kappa_\gamma$  and $\kappa_{Z\gamma}$ from extra Higgs bosons in 2HDM.

The triple couplings among Higgs bosons themselves are relevant for the loop corrections. When omitting the $\mathcal{O}(\cos^2(\beta-\alpha))$, they read
\begin{align}
\label{eq:tri-higgs}
\lambda_{h \Phi\Phi}&=-\frac{C_\Phi  }{2v} \left[m_h^2+2m_\Phi^2-2M^2+2(m_h^2-M^2) \cot 2\beta \cos(\beta-\alpha)\right],  \\
\lambda_{h HH}&=-\frac{1}{2v} \left[m_h^2+2m_H^2-2M^2+2(m_h^2+2m_H^2-3M^2) \cot 2\beta \cos(\beta-\alpha)\right]\,,
\end{align}
with $M^2 \equiv {m_{12}^2}/(\sin\beta \cos\beta)$, $C_\Phi = 2(1)$ for $\Phi = H^\pm (A)$.  
One notable difference between the Type-I study here and our former Type-II study~\cite{Chen:2018shg} is that those terms proportional to $\cosba$ could play a more important role in the Type-I 2HDM. 
In the Type-II 2HDM, Yukawa couplings have both $\cot \beta$ enhancement and $\tanb$ enhancement at the tree level, which tightly constraints the range of $\cos(\beta-\alpha)$ when Higgs precision measurements are considered.  
In the Type-I 2HDM, all Yukawa couplings are proportional to $\cot\beta$ at the leading order, with no large $\tan\beta$ enhancement. 
The viable range of $\cos(\beta-\alpha)$ could be larger when tree-level effects are included. However, given that
 \begin{equation}
\label{eq:loop_tan}
\cot 2\beta \cos(\beta-\alpha) = -\frac{1}{2}\frac{\tan^2 \beta-1}{\tanb} \cosba ,
\end{equation}
loop corrections induced by the triple Higgs couplings $\lambda_{h \Phi\Phi}$, $\lambda_{h HH}$ would have interesting  $\tan \beta$ enhancement that competes with the tree-level corrections, which are usually sub-dominant in Type-II 2HDM, especially in the large $\tan\beta$ region when the tree level effects dominate. 
Once loop effects are included, the range of $\cos(\beta-\alpha)$ at the tree-level loosely constrained large $\tan\beta$ region in the Type-I 2HDM shrinks significantly.

With the degenerate masses of $m_\Phi \equiv m_H = m_A = m_{H^\pm}$ and the  alignment limit of $\cosba=0$, we can introduce a new parameter of $\lambda v^2$ defined as
\begin{equation}
\label{eq:triple}
\lambda v^2 \equiv m_\Phi^2 - \frac{m_{12}^2}{\sin \beta \cos \beta}\,,
\end{equation}
which is the parameter that enters the Higgs self-couplings and relevant for the loop corrections to the SM-like Higgs boson couplings. 
This parameter could be used interchangeably with $m_{12}^2$ as we will do for convenience.  

For the rest of our analysis, we take the input parameters $v=246$ GeV and $m_h=125$ GeV. 
The remaining free parameters are
\begin{equation}\label{eq:para}
\tan\beta\,,\  \cos(\beta-\alpha)\,,\ m_H\,,\ m_A\,,\ m_{H^\pm}\ {\rm and}\ \lambda v^2\,.
\end{equation}
Although these six parameters are independent of each other, their allowed ranges under perturbativity, unitarity, and stability consideration are correlated.

For simplicity, one often begins with the degenerate case where all heavy Higgs boson masses are set the same. We will explore both the degenerate case and deviation from that, the non-degenerate case, specified as 
\beqs\label{eqs:degen}
\begin{eqnarray}
{\rm Degenerate\ Case:\ }&&
m_\Phi \equiv m_H = m_A = m_{H^\pm}\,, \\
{\rm Non\  Degenerate\ Case:\ }&&
\Delta m_{A,C} \equiv m_{A, H^\pm} -m_{H}\,.
\end{eqnarray}
\eeqs
As such, there will be four independent parameters for the degenerate case, and five for the non-degenerate case if assuming $\Delta m_{A} = \Delta m_{C}$. 
With the current LHC Higgs boson measurements~\cite{Coleppa:2013dya, Craig:2013hca,Barger:2013ofa,Belanger:2013xza, ATLAS:2018doi, Sirunyan:2018koj}, deviations of the Higgs boson couplings from the decoupling and alignment limits are still allowed at about $10\%$ level. All tree-level deviations from the SM Higgs boson couplings are parametrized by only two parameters: $\tan\beta$ and $\cos(\beta-\alpha)$. 
Once additional loop corrections are included, dependence on the heavy Higgs boson masses as well as $\lambda v^2$ also enters. 
In our following analyses, we study the combined contributions to the couplings of the SM-like Higgs boson with both tree-level and loop corrections. 
The calculations of $\kappa$'s are performed with full electroweak one-loop corrections,\footnote{\href{https://github.com/ycwu1030/THDMNLO_FA}{https://github.com/ycwu1030/THDMNLO\_FA}.} as discussed in details in Ref.~\cite{Chen:2018shg}.

\section{Theoretical Constraints}
\label{sec:theory}

Heavy Higgs loop corrections will involve the Higgs boson masses and self-couplings, as $\lambda_{1-5}$ in \autoref{eq:L_2HDM}. 
 These parameters are constrained by theoretical considerations such as vacuum stability~\cite{Nie:1998yn}, perturbativity, and partial wave  unitarity~\cite{Ginzburg:2005dt}. 
\begin{itemize}
\item Vacuum Stability
\begin{equation}
\lambda_1>0,\quad \lambda_2>0,\quad \lambda_3>-\sqrt{\lambda_1\lambda_2},\quad \lambda_3+\lambda_4-|\lambda_5|>-\sqrt{\lambda_1\lambda_2}\,.
\end{equation}
\item Perturbativity
\begin{equation}
|\lambda_i|\le 4\pi\,.
\end{equation}
\item Unitarity
\begin{equation}
|a_{i,\pm}^0|\le\frac{1}{2}\,.
\end{equation}
\end{itemize}
The details of $a_{i,\pm}^0$ were shown in Refs.~\cite{Ginzburg:2005dt,Gu:2017ckc}. 
In what follows, we will discuss the constraints in several different cases.

\begin{figure}[h]
\centering
\includegraphics[width=0.45\textwidth]{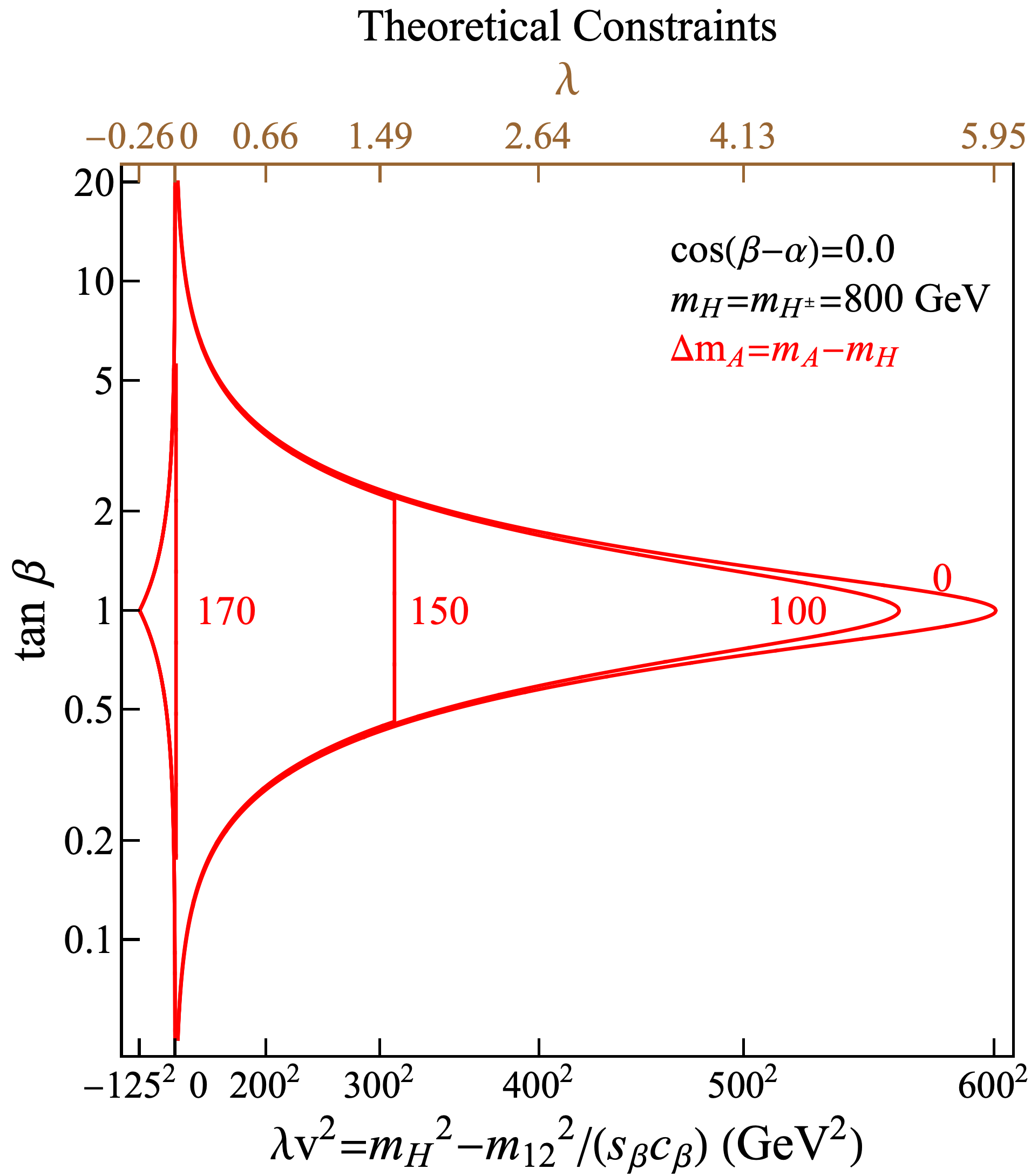}
\includegraphics[width=0.49\textwidth]{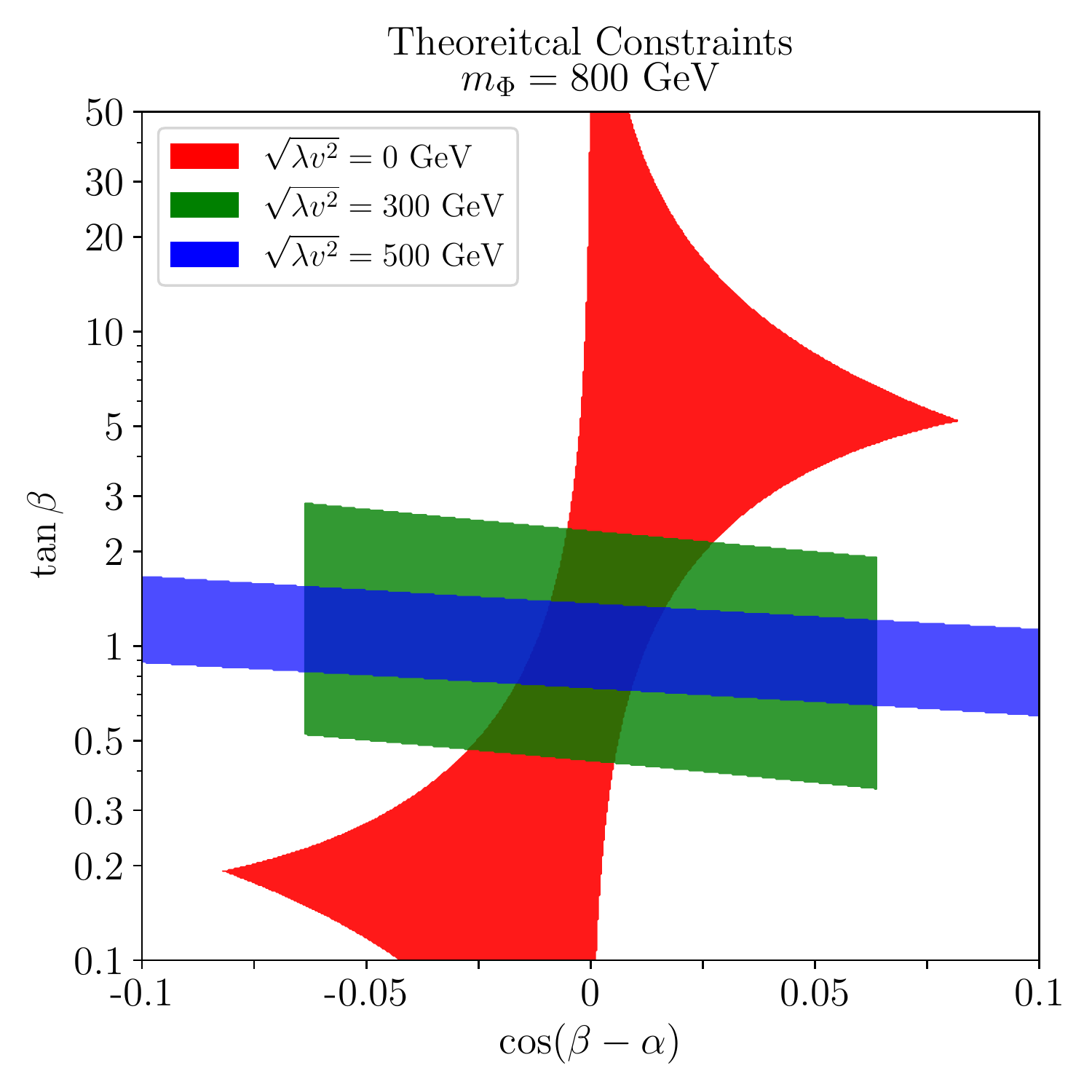}
\caption{Left panel: Allowed regions in the $\lambda v^2-\tan\beta$ plane with all theoretical considerations taken into account, for $m_H=m_{H^\pm}=800$ GeV, with fixed $\cos(\beta-\alpha)=$0 and varying $\Delta m_A=m_{A}-m_H$.
Right panel: Allowed regions in the $\cosba-\tanb$ plane for $m_\Phi$ = 800 GeV, with varying $\lambvs = $ 0 (red), 300 GeV (green), 500  GeV (blue).
}
\label{fig:theory_case2}
\end{figure}

\subsection{Case 1: alignment limit with degenerate heavy Higgs masses}

Theoretical constraints do not depend on the Yukawa structure at the leading order.
Detailed discussions were included in the previous work of~\cite{Gu:2017ckc,Chen:2018shg}. 
In general, $\lambda v^2$ is constrained to be 
\begin{equation}
  -m_h^2 < \lambda v^2 < (600\  \text{GeV})^2\,,
     \label{eq:lam_cons}
\end{equation}
which gives $ -0.258 < \lambda =- \lambda_4=- \lambda_5  < 5.949$ and $0 <  \lambda_3  < 6.207$.  $\tanb$ dependence enters as 
\beqs
\bea
&&\tan^2\beta+\frac{1}{\tan^2\beta}<-\left(\frac{m_h^2}{\lambda v^2}+\frac{\lambda v^2}{m_h^2}\right),\ \ \ {\rm for}\ \lambda v^2 < 0 \,,\label{eq:constn}\\
&&\tan^2 \beta + \frac{1}{\tan^2 \beta} < \frac{64 \pi^2 v^4 + 5 m_h^4 - 48 \pi v^2 m_h^2 + 8 \lambda^2 v^4 - 4 m_h^2 \lambda v^2}{3\lambda v^2(8 \pi v^2 - 3 m_h^2)},\ \ \ {\rm for}\ \lambda v^2 > 0. \ \ \ \label{eq:constp}
\eea
\eeqs
\autoref{eq:constn} mainly comes from the requirement of vacuum stability, while \autoref{eq:constp} is due to the partial wave unitarity. 

In \autoref{fig:theory_case2}, we present contours to illustrate the theoretical constraints on $\tan\beta$ versus the other model parameters as discussed in the beginning of  Sec.~\ref{sec:theory}. As shown in the left panel, for $\lambda v^2=0, m_{H/H^\pm}=800$ GeV, $\tan\beta$ is unconstrained. The allowed range of $\tan\beta$ quickly shrinks as $\lambda v^2$ increases. The degenerate case under consideration for this section is shown by the outer contour $\Delta m_A= m_A-m_{H/H^\pm}=0$.  

\subsection{Case 2: alignment limit with non-degenerate heavy Higgs masses}

For non-degenerate heavy Higgs masses with the mass splittings $\Delta m_{A(C)} = m_{A(H^\pm)} - m_H$, two special cases are of particular interest $m_A=m_{H^\pm}$ and $m_{H^\pm}=m_H$. The theoretical constraints for $m_A=m_{H^\pm}$ have  been discussed in \cite{Chen:2018shg}. Here we focus on the other case  $m_{H^\pm}=m_H$ in which the $Z$-pole constraints are automatically satisfied.

The strongest constraints on parameters are imposed by the partial wave unitarity, in particular,  
$a_{1,+}^0\leq \frac{1}{2}$, which primarily sets limits on the value of $\tan\beta$, and 
$a_{4,-}^0\leq\frac{1}{2}$, which constrains the mass splitting $\Delta m_A=m_A-m_{H/H^\pm}$. The allowed range of $\tan\beta$ for various $\Delta m_A$ are plotted in the left panel of \autoref{fig:theory_case2}, which is  not very sensitive to the mass splitting. It can be well approximated by \autoref{eq:constp}. For large mass splitting, namely $m_A^2-m_H^2\gtrsim\pi v^2$,   $a_{4,-}^0\leq\frac{1}{2}$   sets a strong upper limit on $\lambda v^2$, which is given by
\begin{equation}
2\lambda v^2\leq 8\pi v^2-5(m_A^2-m_H^2)-m_h^2.\label{eq:soluni2}
\end{equation}
This explains the straight right boundary in the left panel of \autoref{fig:theory_case2} as $\Delta m_A\gtrsim 100$ GeV. For $\Delta m_A >0$, the range of $\lambda v^2$ shrinks more for larger $\Delta m_A$: from about ${\rm (600~GeV)}^{2}$ with $\Delta m_A=0$ to about ${\rm (300~GeV)}^{2}$ for $\Delta m_A=150$ GeV.

%
 
\subsection{Case 3: non-alignment limit with degenerate heavy Higgs masses}
 
The theoretical constraints also limit the range of $\cos(\beta-\alpha)$, as shown in the right panel of \autoref{fig:theory_case2} for the degenerate case with different values of $\lambda v^2$ for  $m_\Phi =800$ GeV.  A larger value of $\lambda v^2$ leads to a more relaxed range of $\cos(\beta-\alpha)$, but a stronger constraint on $\tan\beta$.  A larger value of $m_\Phi$ leads to a smaller region in $\cos(\beta-\alpha)$.  One interesting feature is a symmetry: for $\cos(\beta-\alpha)\to -\cos(\beta-\alpha)$, $\tan\beta\to \cot\beta$, which is evident in the right panel of \autoref{fig:theory_case2} as well.

\section{Current and Expected LHC Search Bounds}
\label{sec:LHC}

The heavy Higgs bosons in the 2HDM have been searched for at the LHC Run-I and Run-II via various channels.
The direct searches include the decay channels of $\tau\tau$~\cite{Aaboud:2017sjh, CMS-PAS-HIG-17-020},    $\mu\mu$~\cite{Sirunyan:2019tkw,Aaboud:2019sgt}, $b\bar b$~\cite{Sirunyan:2018taj, CMS:2019hvr,Aad:2019zwb}, $WW/ZZ$~\cite{Aaboud:2017gsl,Aaboud:2017rel,Sirunyan:2018qlb}, $\gamma\gamma$~\cite{Aaboud:2017yyg}, $H\to hh$~\cite{Aaboud:2018ftw, Sirunyan:2018iwt}, $A\to hZ$~\cite{Aaboud:2017cxo}, and $A/H\to HZ/AZ$~\cite{Aaboud:2018eoy,Khachatryan:2016are}.  Since the limits from heavy Higgs decays to $\mu\mu$ and $b\bar b$ are always similar but weaker comparing to the $\tau\tau$ channel, we will only show constraints from $\tau\tau$ channel here. In addition, the leading decay modes of heavy neutral Higgs bosons is $A/H\to t \bar t$, which was known to have strong signal-background interference effects~\cite{Dicus:1994bm}. 
Recent studies of the LHC 8 TeV search sensitivities via this channel can be found in Refs.~\cite{Aaboud:2017hnm}, utilize the lineshape of $t\bar{t}$ invariant mass distribution~\cite{Djouadi:2015jea,Carena:2016npr,Djouadi:2019cbm}, and the experimental searches were made in Ref.~\cite{Sirunyan:2019wph}.
Knowing the search limit at $\sqrt{s} =8$ TeV or $\sqrt{s}=13$ TeV, the associated limit at $\sqrt{s}=14$ TeV could be estimated through a  scaling relation \cite{Djouadi:2015jea}
\begin{eqnarray}
R_{14}^S(M_{A/H} )&\approx& \sqrt{{\cal L}_{8/13} /{\cal L}_{14} } \times \sqrt{\sigma_{14}^S /\sigma_{8/13}^S } \times R_{8/13}^S(M_{A/H})\,.
\end{eqnarray}

To use the published cross-section times branching ratio limits to directly constrain the 2HDM parameter space, we work with the {\tt SusHi} package \cite{Liebler:2016ceh} for the production cross-section at the NNLO level, and our own improved {\tt 2HDMC} code, which adds loop-level effects to the public {\tt 2HDMC} code~\cite{Eriksson:2009ws}, for the branching ratios.
\begin{figure}[h]
\centering
\includegraphics[width=0.48\textwidth]{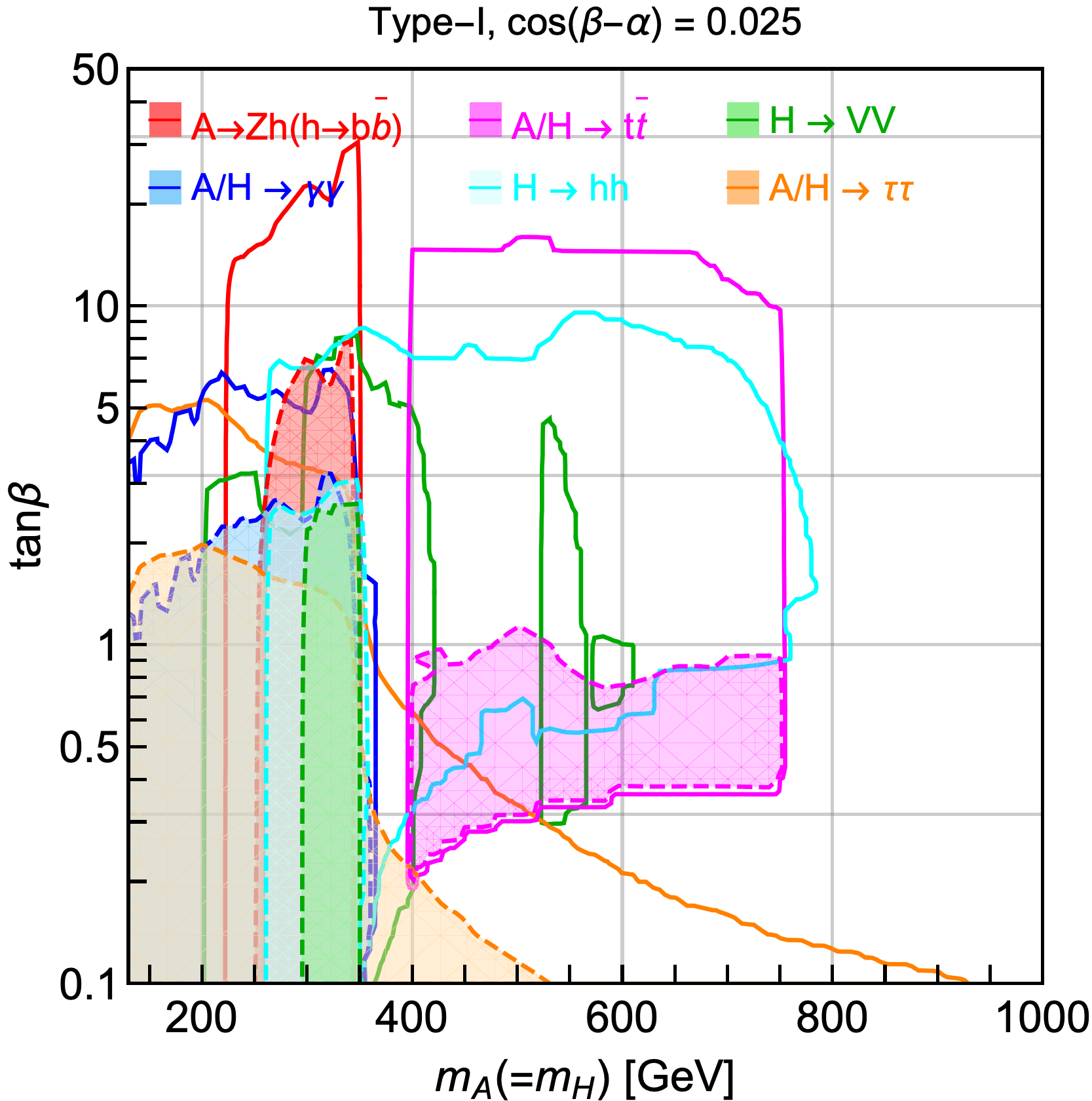}
\includegraphics[width=0.48\textwidth]{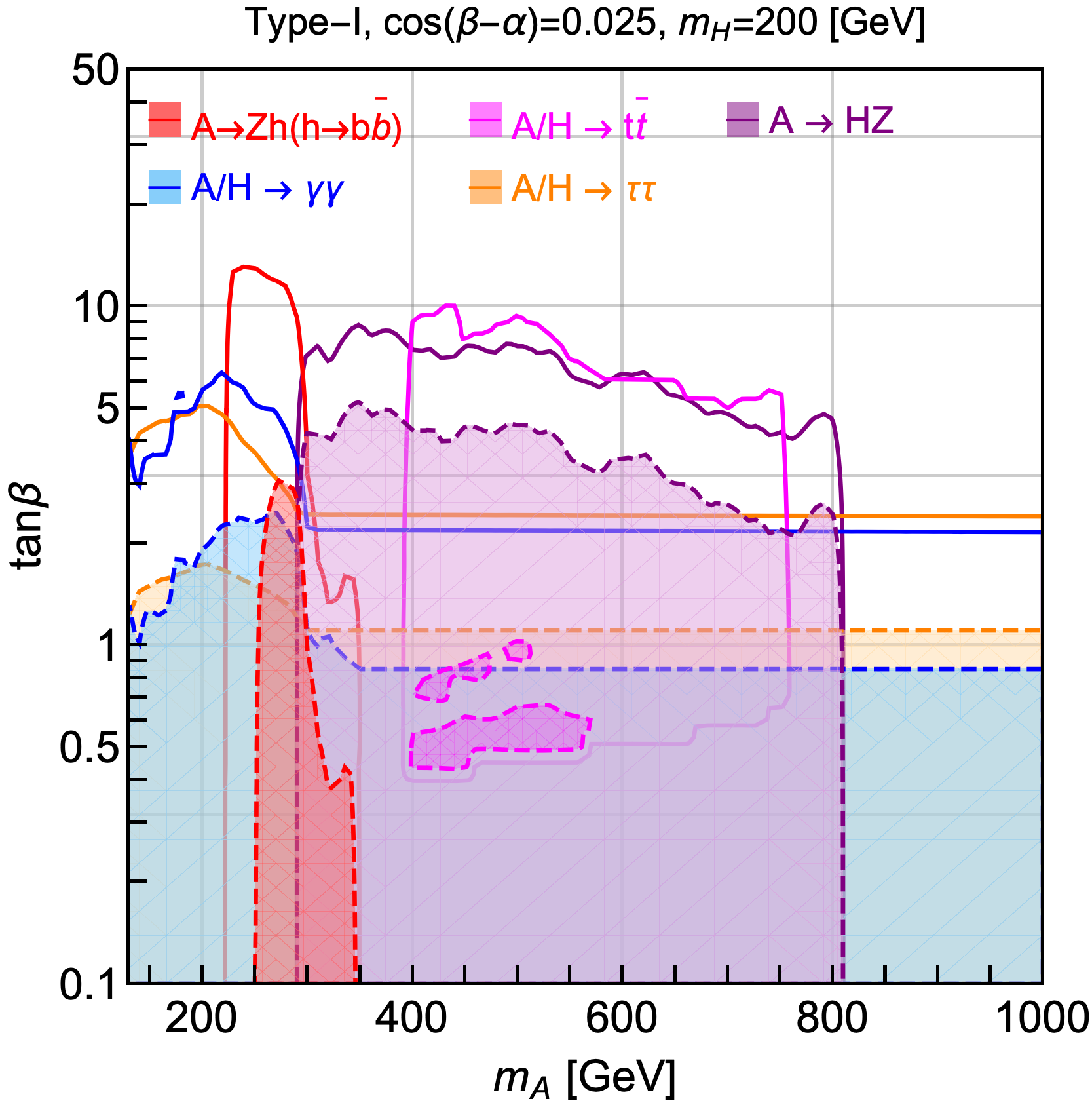}
\caption{Excluded regions at 95\% C.L. in the $m_{A}-\tan\beta$ plane from the LHC Run-II (regions with dashed line boundaries) and HL-LHC (solid lines) via different channels: $\tau\tau$ (orange), $t \bar t$ (magenta), $VV$ (green), $\gamma\gamma$ (blue), $H\to hh$ (cyan), $A\to hZ$ (red), and $A\to HZ$ (purple).  The left panel is for the degenerate case, with benchmark parameter $\cosba=0.025$, while the right panel is for the non-degenerate case with fixed $m_H=200$ GeV  and $\cosba=0.025$.
  }
\label{fig:TypeIAHExcl}
\end{figure}

 In the left panel of~\autoref{fig:TypeIAHExcl}, we present the 95\% C.L. limits of the neutral Higgs boson searches in the $m_{A/H} - \tan\beta$ plane for the degenerate case under the current LHC Run-II searches (shaded region enclosed by the dashed lines) as well as the projected HL-LHC search limits (region enclosed by the solid lines).  We have chosen the  non-alignment case of $\cos(\beta-\alpha)=0.025$, in which $H \to hh$, $A\rightarrow hZ$,  and $WW/ZZ$ channel contribute.  Unlike the Type-II 2HDM case in which there are very strong experimental search limits from $\tau\tau$ channel at large $\tanb$, for the Type-I 2HDM, large $\tan\beta$ region is basically unconstrained since all the Yukawa couplings are proportional to $1/\tan\beta$. For the low $\tan\beta\sim 0.1$ region, $\tau\tau$ mode has the best reach: the current LHC Run II excludes heavy Higgs mass up to about 550 GeV, and the exclusion reach is about 950 GeV at HL-LHC.  $\gamma\gamma$, $VV$, $Zh$, and $hh$  channels exclude $\tan\beta$ up to about 5 for $m_{A/H}<2 m_t$ under the current LHC Run-II, and up to about 30 at HL-LHC.  For $m_{A/H}>2 m_t$, $tt$ mode provides the best reach: $\tan\beta$ is excluded up to about 1 at the current LHC Run-II, and up to about 15 at the HL-LHC, for $m_{A/H}\lesssim 750$ GeV.
Comparing to the exclusion under the alignment limit in which $Zh$, $hh$ and $VV$ channels are absent, the limits of the  $\tau\tau$, $\gamma\gamma$  channel are relaxed slightly given the opening of $H \to hh$, $WW/ZZ$, and  $A\rightarrow hZ$.  

In the right panel of~\autoref{fig:TypeIAHExcl}, we show the exclusion region in the $m_A - \tanb$ plane for $m_H=200$ GeV and $\cba=0.025$.    Additional exotic decay channel of $A \rightarrow HZ$ contributes, which is shown in the purple shaded region.  It covers the entire mass region of 350 GeV $< m_A < 800$ GeV for $\tan\beta\lesssim 5 (10)$ region for the current LHC Run-II (HL-LHC).
Low $\tan\beta\lesssim 1 (2)$ region for the current LHC Run-II (HL-LHC) is excluded by 200 GeV $H\rightarrow \gamma\gamma$ and $\tau\tau$.  $A\rightarrow Zh$ and $tt$ channels are still effective, with relaxed limits comparing to the degenerate case, given the opening of $A \rightarrow HZ$.

\begin{figure}[h]
\centering
\includegraphics[width=0.48\textwidth]{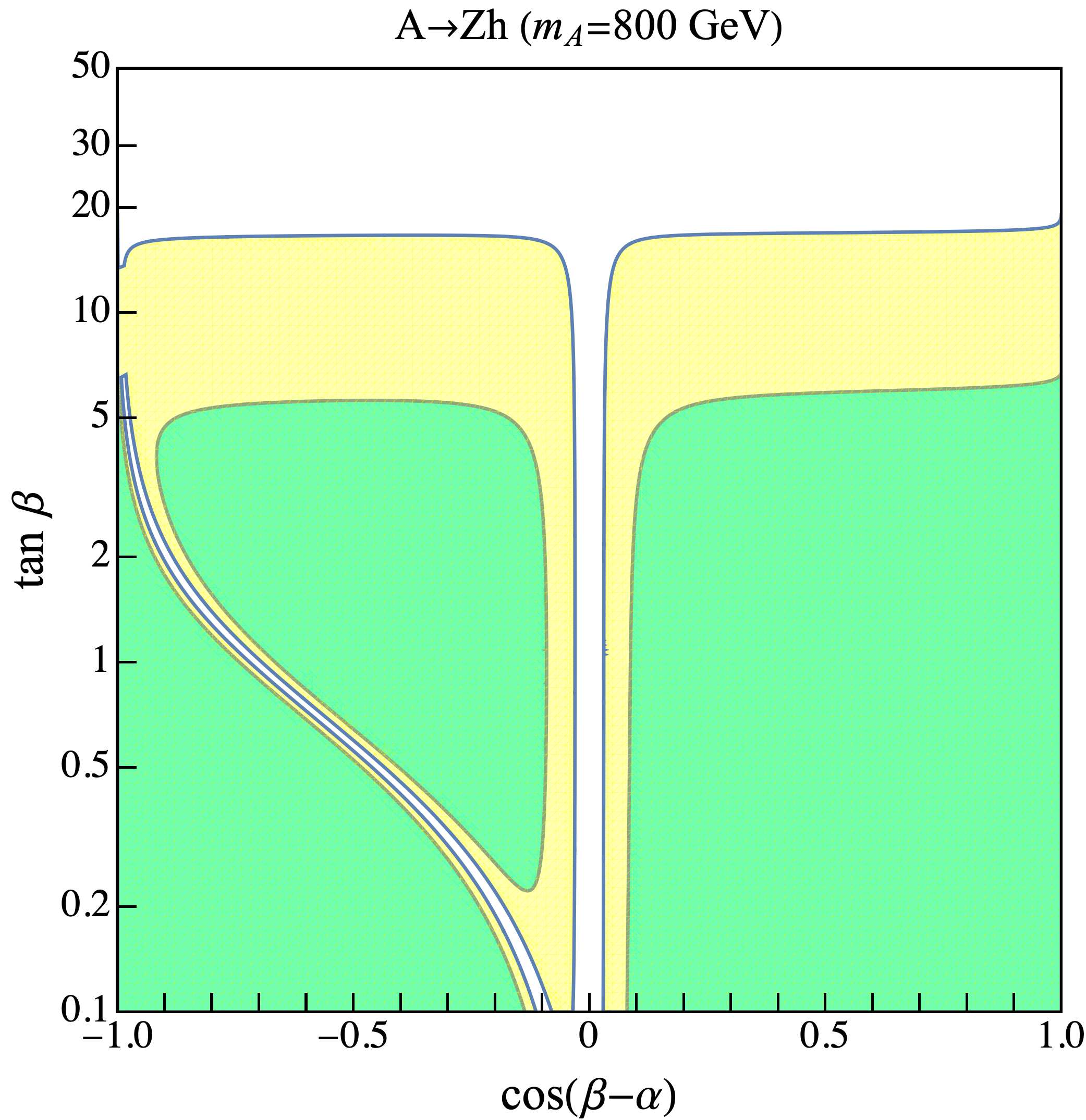}
\includegraphics[width=0.48\textwidth]{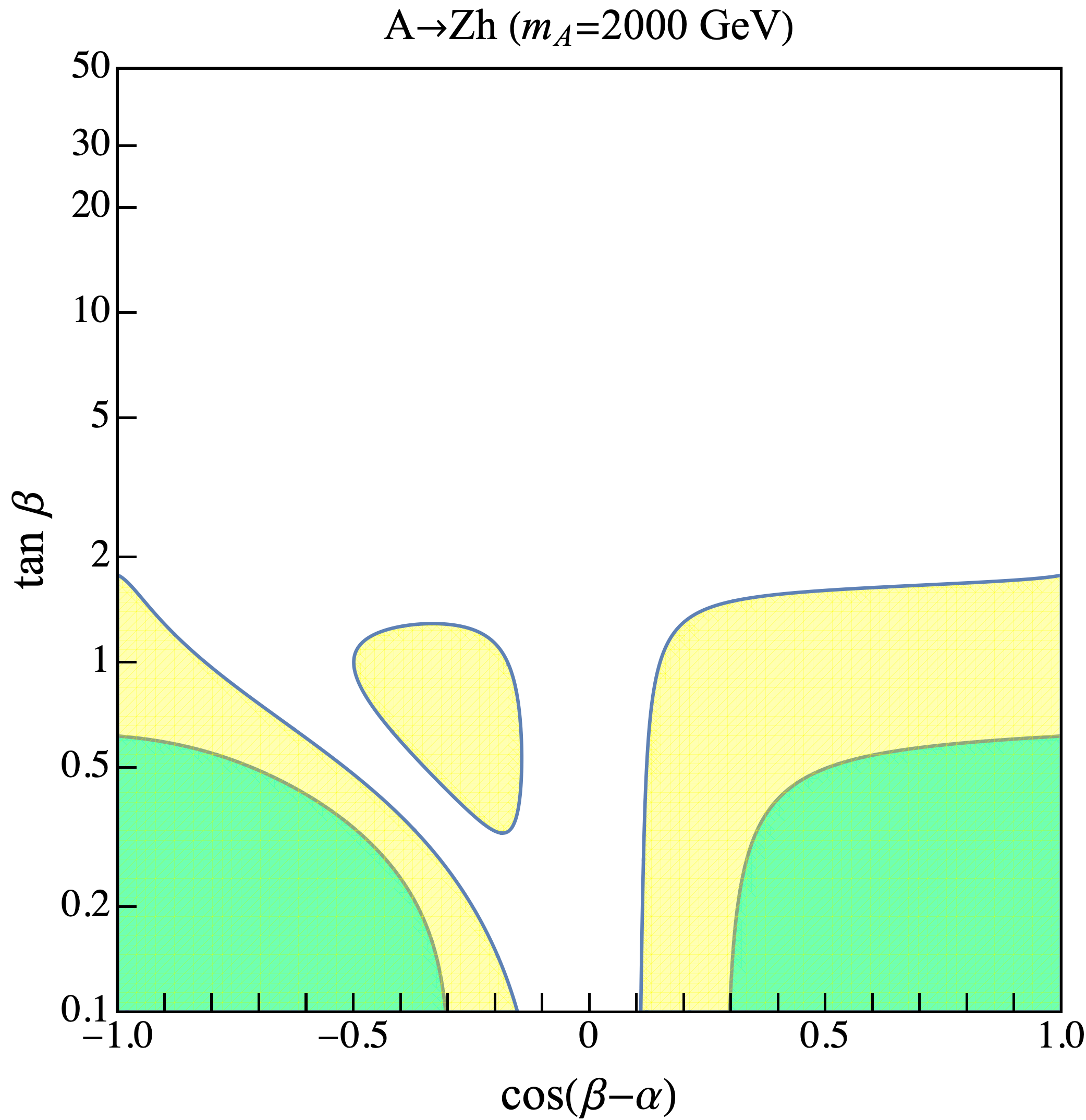}
\caption{Excluded regions at 95\% C.L. in the $\cos(\beta-\alpha)-\tan\beta$ plane from the LHC direct searches for the CP-odd heavy Higgs boson via the $A\to hZ$ decays, with $m_A=800$ GeV (left panel) and $m_A = 2000$ GeV (right panel).
The shaded regions are excluded under the current LHC 13 TeV $36.1$ fb$^{-1}$ (green) and the future projected HL-LHC 14 TeV $3000$ fb$^{-1}$ (yellow), respectively.    
}
\label{fig:TypeIAhZ}
\end{figure}

In Fig.~\ref{fig:TypeIAhZ}, we present the 95\% C.L. excluded region in the  $\cos(\beta-\alpha) - \tan\beta$ plane for the LHC $13$ TeV searches \cite{Aaboud:2017cxo} and for the projected HL-LHC $14$ TeV sensitivity via the $A\to hZ$ channel. 
The results were shown for two fixed heavy CP-odd Higgs boson masses of $m_A=800$ GeV and $m_A=2000$ GeV, respectively.  
For the case of $m_A=800$ GeV, a narrow band for $\cos(\beta-\alpha) \lesssim -0.1$ or $| \cos(\beta-\alpha) | \lesssim 0.02$ is allowed by  the HL-LHC data.
In addition, the large-$\tan\beta$ regions of $\tan\beta\gtrsim 5$ and $\tan\beta\gtrsim 20$ are also allowed at the current LHC and the future HL-LHC searches,   which is due to the suppressed production cross sections of $\sigma(gg\to A)$ and $\sigma(b \bar b \to A)$ in the Type-I 2HDM case.
For the $m_A=2000$ GeV case, the current and the future LHC searches for the $A\to hZ$ can only exclude the regions with small input values of $\tan\beta$,  due to the suppressed production cross section for heavy Higgs bosons.  Note that the strong constraints usually present in the Type-II case at large $\tan\beta$ is again absent here in the Type-I case, due to the $1/\tan\beta$ dependence of the Higgs Yukawa couplings.

\section{Fitting Results}
\label{sec:results}

\subsection{$\chi^2$ fit framework}

With the Higgs precision measurements summarized in Table.~\ref{tab:mu_precision}, we performed a $\chi^2$ fit to determine the allowed parameter space of Type-I 2HDM.  
With the same method described in Ref.~\cite{Chen:2018shg}, we construct  the $\chi^2$ with the profile likelihood method,
\begin{equation}
\chi^2=\sum_i\frac{(\mu_i^{\rm{BSM}}-\mu_i^{\rm{obs}})^2}{\sigma_{\mu_i}^2}\,,
\end{equation}
where $\mu_i^{\rm{BSM}}=(\sigma\times\textrm{Br}_i)_{\rm{BSM}}/(\sigma\times\textrm{Br}_i)_{\rm{SM}}$ is the signal strength for various Higgs search channels, $\sigma_{\mu_i}$ is the estimated error for each process. 
Usually, the correlations among different $\sigma\times\rm{Br}$ are not provided and are thus assumed to be zero. 
For future colliders, $\mu_i^{\rm{obs}}$ are set to be unity in the current analyses, assuming no deviations from the SM observables.\footnote{If deviations are observed in the future, we can use the same $\chi^2$ fit method to determined the constrained parameter space, with $\mu_i^{\rm{obs}}$ being the observed experimental central value.  Detailed work along this line is currently under study~\cite{2019NTLSSW}.} In this work, we will focus specifically on CEPC, and also compare the reaches of three future lepton colliders, including ILC and FCC-ee.

The overall $\chi^2$ is calculated by substituting the $\kappa$'s defined in~\autoref{eq:Leff} into corresponding $\mu_i^{\rm BSM}$. In the rest of the analyses, we determine the allowed parameter region at the $95\%$ Confidence Level (C.L.).  For the one-, two- or three-parameter fit, the corresponding $\Delta\chi^2=\chi^2-\chi_{\rm{min}}^2$ at 95\% C.L. is 3.84, 5.99 or 7.82, respectively. Note that when we present our results with three-parameter fit, we project the three-dimension space into two-dimension plot, and choose several benchmark points in the third dimension of the parameter space for illustration. 

\subsection{Case with Degenerate Heavy Higgs Masses}

We will first show our results in the case with degenerate heavy Higgs masses,  $m_\Phi=m_H=m_A=m_{H^\pm}$,  which satisfies the $Z$-pole physics constraints automatically.

\subsubsection{Constraints in the $\cos(\beta-\alpha)-\tan\beta$ plane}

\begin{figure}[h]
\centering
\includegraphics[width=0.6\textwidth]{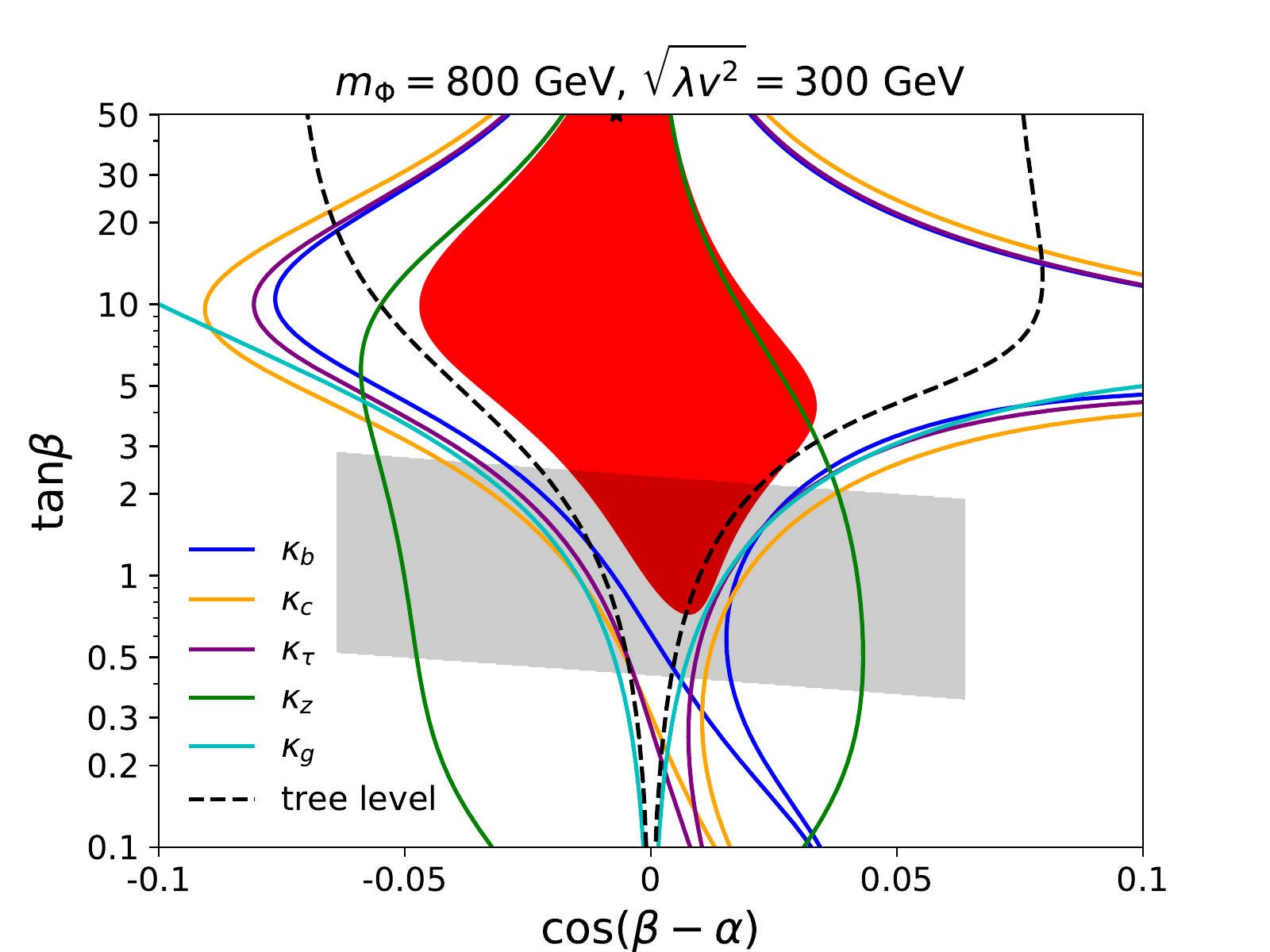}
\caption{\small Allowed region from  two-parameter fitting results at 95\% C.L. in the $\cos(\beta-\alpha)-\tan\beta$ plane under CEPC Higgs precision measurements at one-loop level for $m_\Phi =$ 800 GeV and $\sqrt{\lambda v^2} =$ 300 GeV. 
The red region is the $\chi^2$ fit result with the best fit point indicated by the black star (near $\tanb=50$ and $\cosba=0$). As a comparison, the black dashed line shows the allowed region at tree level.  
Regions enclosed by curves of different colors indicate the domains allowed by individual coupling measurements. The grey shadow area indicates the theoretically favored region.}
\label{fig:tanbcba_fit_degenerate}
\end{figure}

For the case with degenerate masses, we first show the result in \autoref{fig:tanbcba_fit_degenerate} of the two-parameter $\chi^2$ fit in the  $\cos(\beta-\alpha)-\tan\beta$ plane at 1-loop level for $m_\Phi=800$ GeV and $\sqrt{\lambda v^2} =$ 300 GeV. 
The red region represents the overall allowed region with the CEPC precision measurements at 95\% C.L. at one-loop level, while the black dashed line represents the allowed region at tree level.  
Individual constraints from $hbb$, $hcc$, $h\tau\tau$, $hZZ$ and $hgg$ are also shown by colored solid lines.
The constraints from $hWW$  and $h\gamma\gamma$ are much weaker due to worse experimental precisions, hence they are not shown   in~\autoref{fig:tanbcba_fit_degenerate}.

Compared with the tree-level dashed line, the allowed red region at the loop level has quite different behaviors at both large and small $\tanb$ regions.  
For small $\tan\beta$, the overall allowed region is mainly constrained by $\kappa_f$ due to the large $\cot\beta$ enhancement of fermion Yukawa couplings at the tree level, while the constraints from $\kappa_Z$ is weak despite its high precision. This makes the outline of red region close to that of the tree level. 
However, the individual fermion lines show peculiar distortion away from the tree level result.
Such modification is related to $\sqrt{\lambda v^2}$ term in the  triple Higgs couplings $\lambda_{h\phi\phi}$, and Yukawa couplings. The effect is proportional to $\cot^2\beta$,  therefore more pronounced when $\tan\beta$ is small.  In particular, the $hbb$ coupling plays an important role in the distortion from the tree-level results due to the large top Yukawa coupling in the top loop contributions.

In the meantime, all couplings at one-loop level significantly deviate from those of tree level ones at large $\tan\beta$. For the Type-I 2HDM there is no $\tanb$ enhancement at tree level, and the main constraint at large $\tanb$ region comes from the precise $hZZ$ coupling measurement. At one-loop level,  the strongest constraint is still from $hZZ$ coupling, because all the SM-Higgs couplings receive a universal $\tanb$ enhancement from the Higgs field renormalization
\begin{align}
\label{eq:simple_kb}
\kappa^{\rm 1-loop}-\kappa^{\rm tree} &\propto \kappa^{\rm tree} \left(\frac{\lambda_{hH^+H^-}^2}{ m_{H^\pm}^2} + \frac{2\lambda_{hAA}^2}{ m_A^2} + \frac{2\lambda_{hHH}^2}{ m_H^2}\right) \propto \kappa^{\rm tree}\frac{m_\Phi^2\tan^2\beta}{  v^2}\cos^2(\beta-\alpha).
\end{align}
Some useful formulas for the analysis are given in \autoref{sec:app}.  The allowed region of $\cos(\beta-\alpha)$ is greatly reduced, comparing to the tree-level results, as shown in~\autoref{fig:tanbcba_fit_degenerate}.

\begin{figure}[h]
  \centering
    \includegraphics[width=0.48\linewidth]{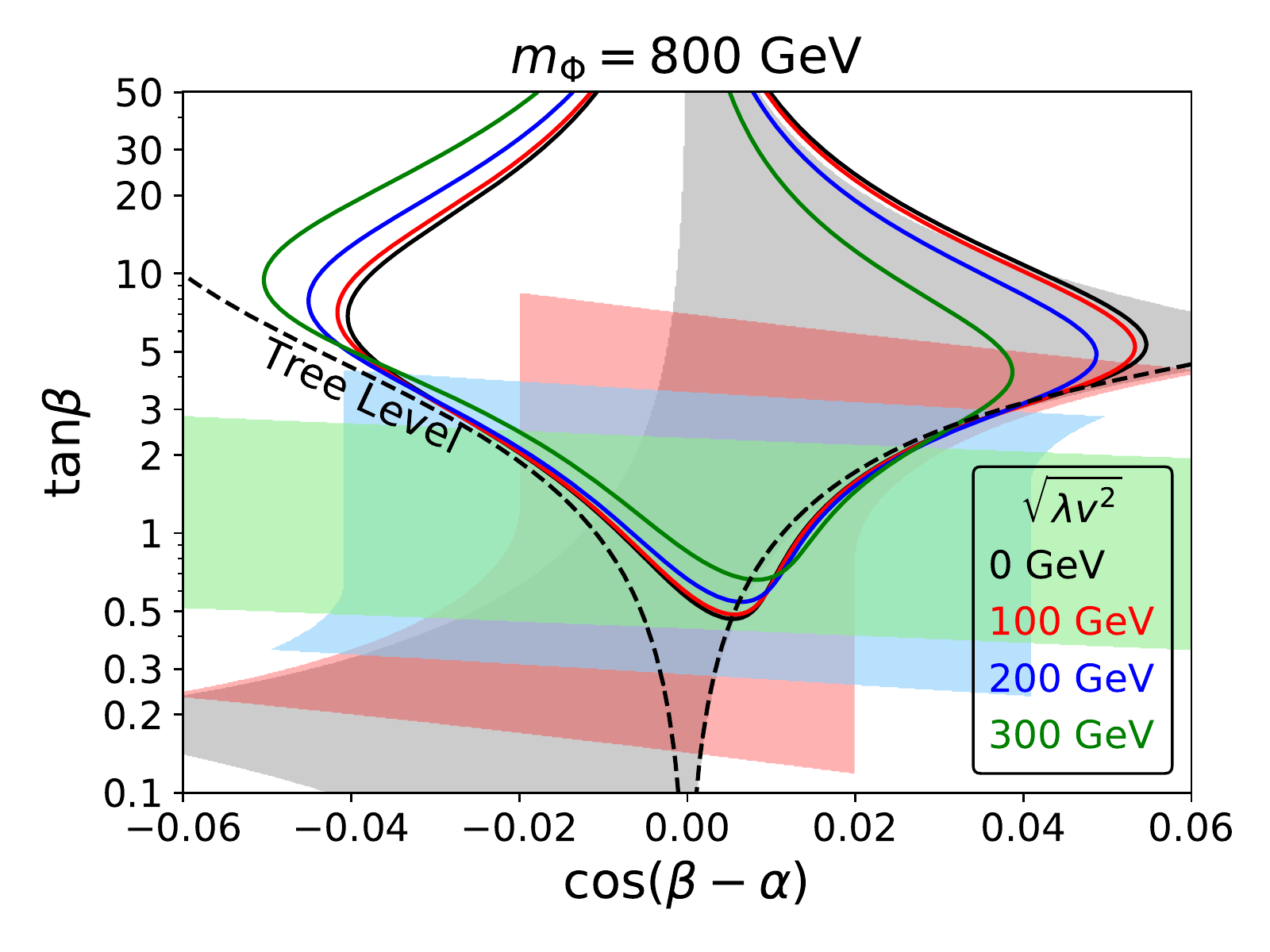}
    \includegraphics[width=0.48\linewidth]{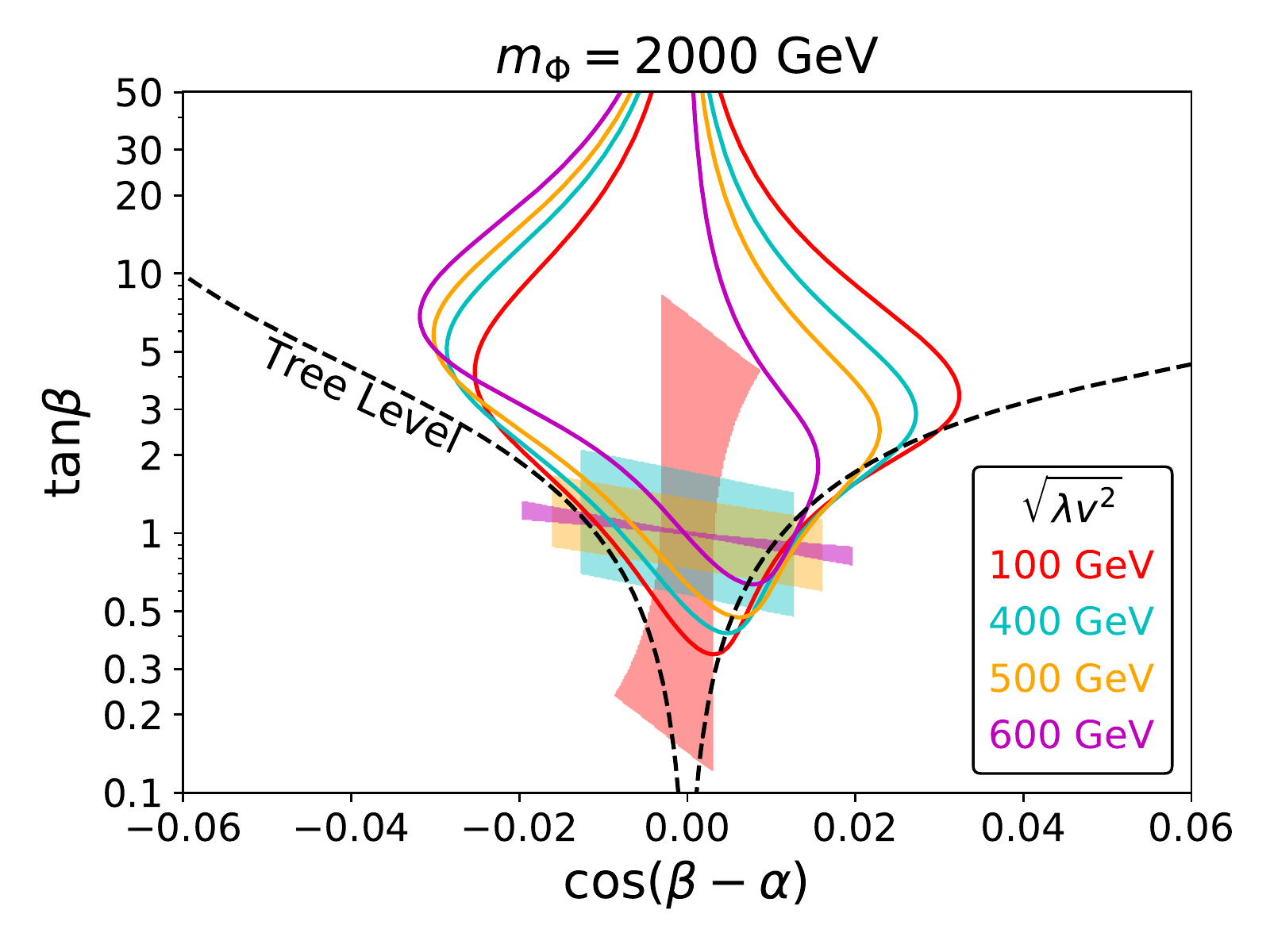}
  \caption{\small  Allowed region from  three-parameter fitting results at 95\% C.L. in the 
  $\cos(\beta-\alpha)-\tan\beta$ plane with varying $\sqrt{\lambda v^2}$ under CEPC precision.   $m_\Phi$ is set to be 800 GeV (left) and 2000 GeV (right). For each $\lambvs$, we show the $\chi^2$ fit result with colored solid lines and the same color shaded region preferred by theoretical constraints. As a comparison, tree-level $\chi^2$ fitting result is shown by dashed black line.}
  \label{fig:tanb_cba_sum}
\end{figure}

We perform a three-parameter fit for $\cos(\alpha-\beta)$, $\tan\beta$,  and $\sqrt{\lambda v^2}$.  Owing to the one more free parameter in the fit and the correlation among the parameters, the allowed region could be different   from that of  the two-parameter fit. 
Figure \ref{fig:tanb_cba_sum} shows the fitting results for a fixed value $m_\Phi$ = 800 GeV with various values $\lambvs$ = 0, 100, 200, 300 GeV (left panel) and $m_\Phi$ = 2000 GeV with $\lambvs$ = 100, 400, 500, 600 GeV (right panel), indicated by different colored lines. With three-parameter fit, $\lambvs \geq 400$ GeV for $m_\Phi=800$ GeV is excluded.
In general, including loop corrections shrinks the allowed parameter space, especially for large and small $\tanb$.  For larger $\lambvs$, there will be a larger asymmetry with respect to $\cosba=0$. The asymmetry at small $\tanb$ is the result from the loop-level $hbb$ couplings because of the large top Yukawa contribution. At large $\tanb$, on the other hand, it is from $hZZ$ coupling, because the triple Higgs couplings as in \autoref{eq:tri-higgs} have terms proportional to $\cosba$.
There is also no decoupling effect for regions with non-zero $\cosba$: the allowed region is smaller with larger $m_\Phi$. This is because a  large part of regions is outside of the theoretically allowed region, shown by the colored shaded region in \autoref{fig:tanb_cba_sum}. 
 
\subsubsection{Constraints on Heavy Scalar Masses}
\begin{figure}[h]
  \centering
    \includegraphics[width=0.48\textwidth]{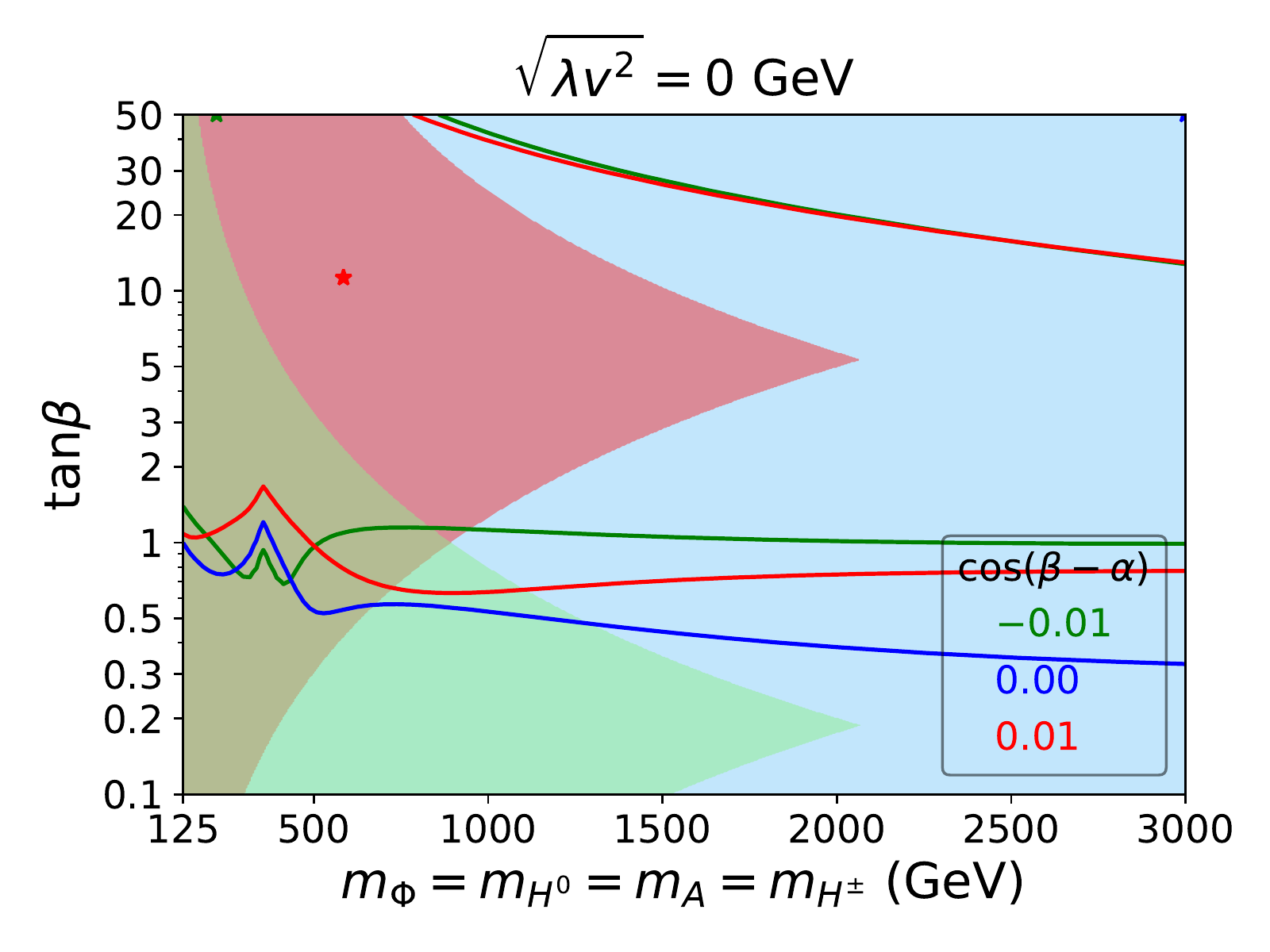}
    \includegraphics[width=0.48\textwidth]{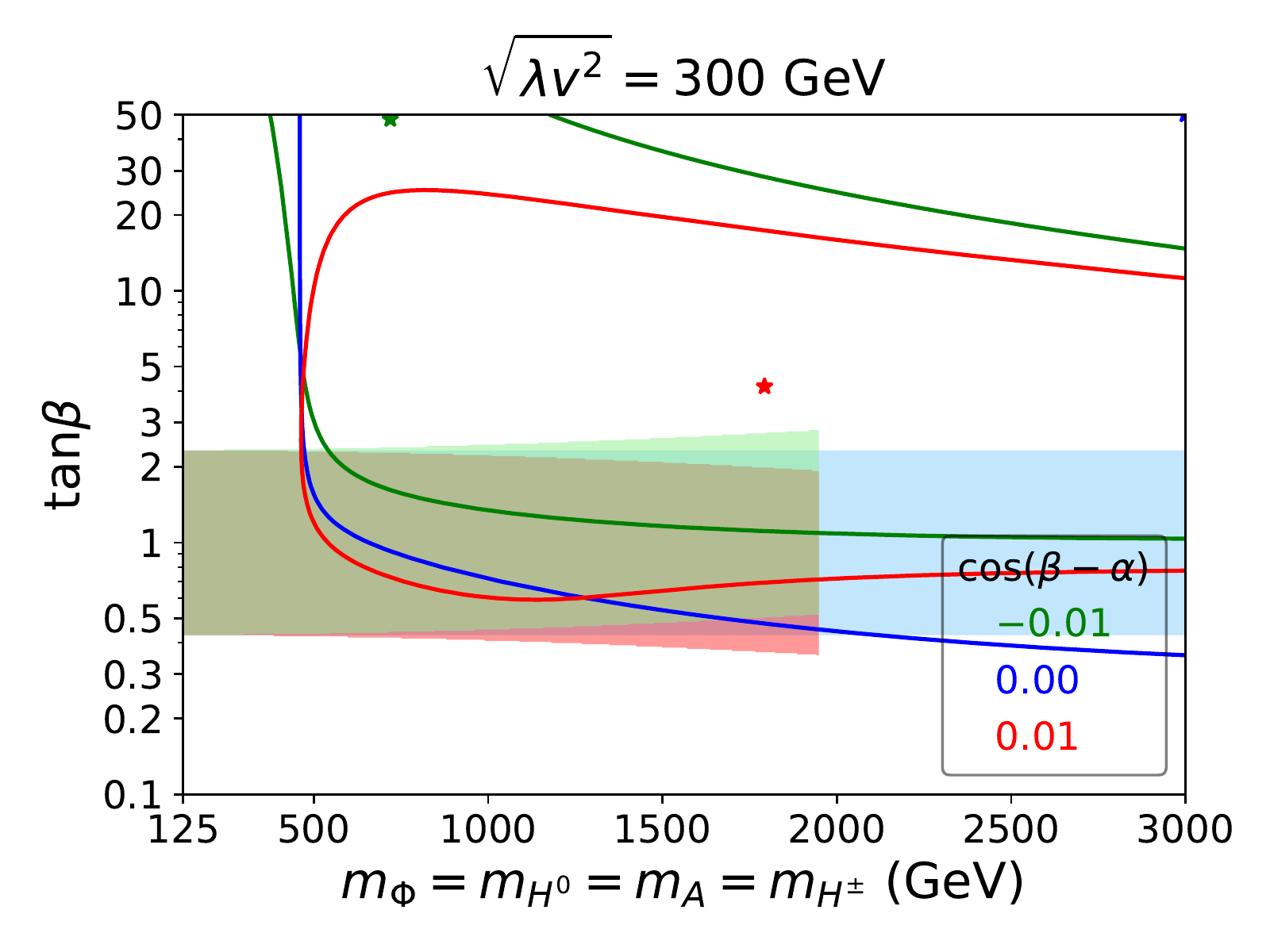}
  \caption{\small Allowed region from  three-parameter fitting results at 95\% C.L. in the $m_\Phi-\tan\beta$ plane with varying $\cosba$ under CEPC precision for  $\sqrt{\lambda v^2}=0$ GeV (left panel) and 300 GeV (right panel) for degenerate mass case. Green, blue and red curves (stars) represent the constraints (best fit point) for  $\cos(\beta-\alpha)=-0.01,\ 0,\ 0.01$ respectively. The theoretical allowed region are also shown in shaded region with the same color.}
  \label{fig:tbvsmass}
\end{figure}

To see how precision measurement constrains heavy scalar masses, we also explore the $\chi^2$ fit in three parameters by fixing $\lambda v^2$ (or $m_{12}^2$). 
Figure \ref{fig:tbvsmass} shows the fitting results at 95\% C.L.~in the $m_\Phi - \tan\beta$ plane under CEPC precision for $\sqrt{\lambda v^2}=0$ GeV (left panel) and 300 GeV (right panel) for the degenerate mass case. Green, blue and red curves (stars) represent the constraints (best fit point) for $\cos(\beta-\alpha)=-0.01,\ 0,\ 0.01$ respectively. The theoretical allowed regions are also shown in the shaded areas with the same color.

For $\sqrt{\lambda v^2}=0$, the constraint on the heavy Higgs mass is rather loose.  Once $\tan\beta\gtrsim 1$, all mass values greater than $m_h$ are allowed.  For nonzero $\cos(\beta-\alpha)$, large $\tan\beta$ are excluded. 
For $\sqrt{\lambda v^2} = 300$ GeV, the heavy Higgs mass is constrained to be larger than about 500 GeV.  For $\cos(\beta-\alpha)=0.01$,  $\tan\beta$ is constrained to be in the range of 0.5 and 20 at 95\% C.L., while for $\cos(\beta-\alpha)=-0.01$, larger values of $\tan\beta$ is allowed.

\begin{figure}[h]
\centering
    \includegraphics[width=0.48\textwidth]{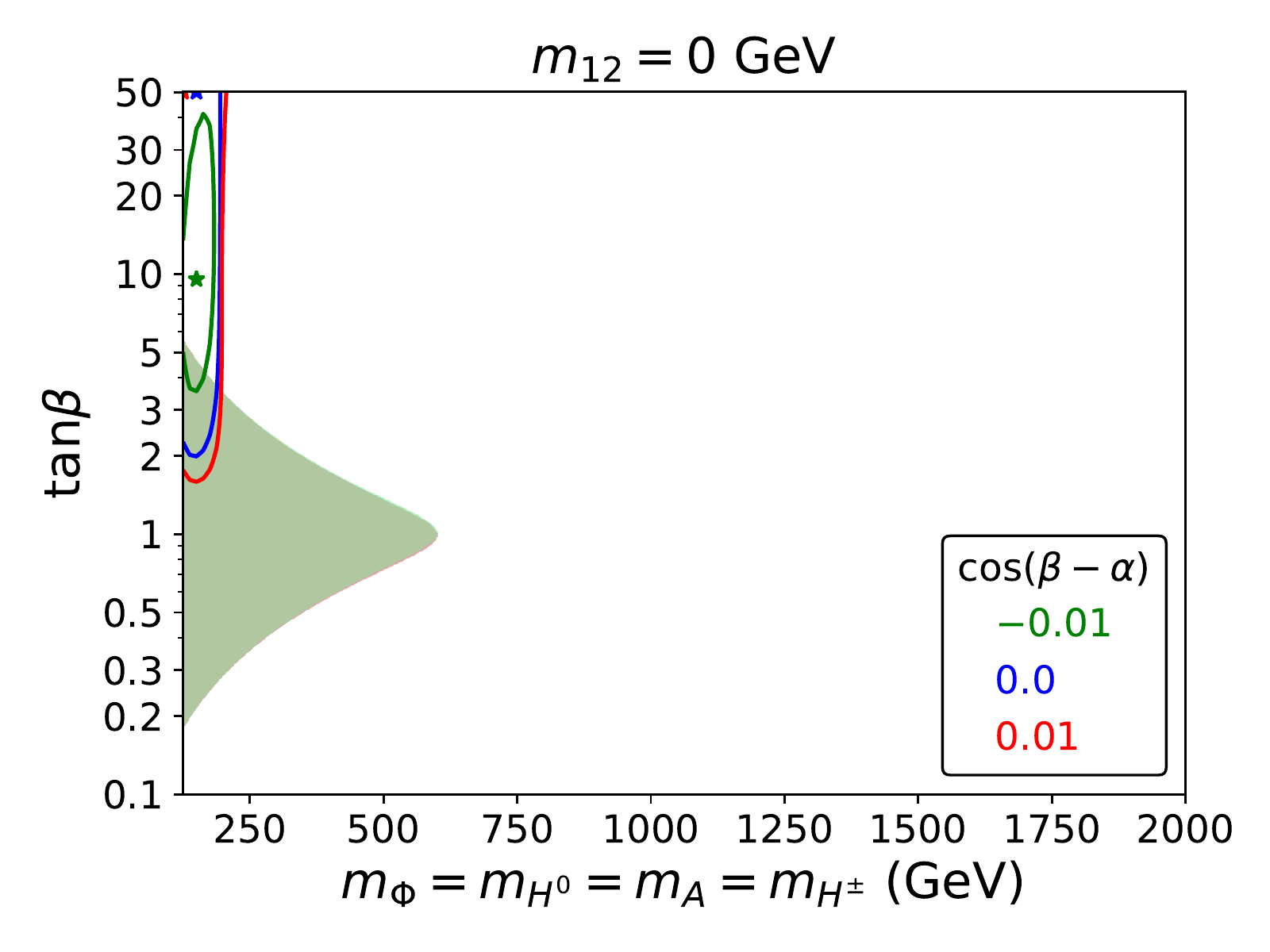}
    \includegraphics[width=0.48\textwidth]{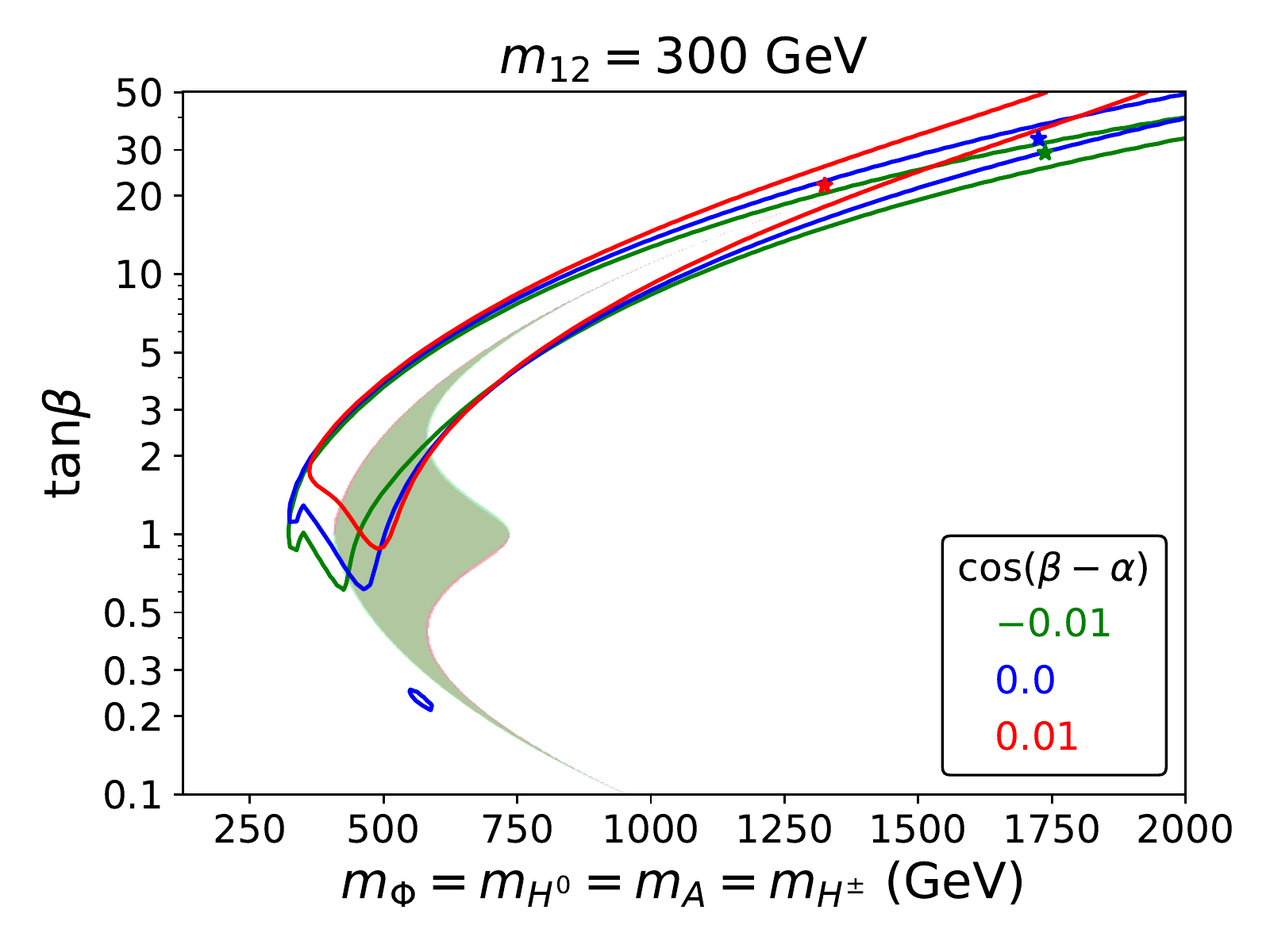}
  \caption{\small Allowed region from  three-parameter fitting results at 95\% C.L. in the $m_\Phi-\tan\beta$ plane with varying $\cosba$ under CEPC precision for $m_{12}=0$ GeV (left panel) and 300 GeV (right panel). The color codes are the same as~\autoref{fig:tbvsmass}.}
  \label{fig:tbvsmassm12}
\end{figure}

It is also interesting to see how future precision measurements constrain the soft $Z_2$ breaking parameter $m_{12}^2$. Figure \ref{fig:tbvsmassm12} shows the fitting results similar to~\autoref{fig:tbvsmass} but for fixing value of $m_{12}$ instead of fixing $\lambda v^2$.  For $m_{12}=0$, $m_\Phi = \sqrt{\lambda v^2}$ is constrained to be less than around 250 GeV.   
For larger values of  $m_{12}$, the rather narrow region in the plane as seen in the right panel indicates a strong correlation between $m_\Phi$ and $\tan\beta$, approximately scaled as $\tan\beta \sim (m_\Phi / m_{12})^2$, which minimizes the corresponding $\lambda v^2$  value and thus its loop effects.

Comparing with the expected direct search limits of heavy Higgs bosons at the HL-LHC, as shown in  \autoref{sec:LHC}, we see that the indirect sensitivities of the SM-like Higgs precision measurements to the heavy Higgs masses and values of $\tanb$ obtained here complement the direct search very well.  While the direct searches usually have better reach in the mass of the heavy Higgs bosons, its sensitivity reduced greatly for large $\tan\beta$ given the suppressed Yukawa couplings.   The indirect reach, on the other hand, tightly constraints the large $\tan\beta$ region when away from the alignment limit, given the enhanced tri-Higgs couplings which enter the corrections to the SM-like Higgs couplings at the loop level.
 
\subsection{Case with Non-degenerate Heavy Higgs Masses}
 
In this section, we go beyond the mass degenerate case and investigate how the Higgs coupling precision measurements could constrain the mass splittings among $m_H$, $m_A$ and $m_{H^\pm}$, and how it complements the $Z$-pole precision measurements.  For $Z$-pole precision measurements, we fit for  the oblique parameters $S$, $T$ and $U$, including the correlation between those oblique parameters, as described in detail in~\cite{Chen:2018shg}.

\begin{figure}[h]
\centering
    \includegraphics[width=0.48\linewidth]{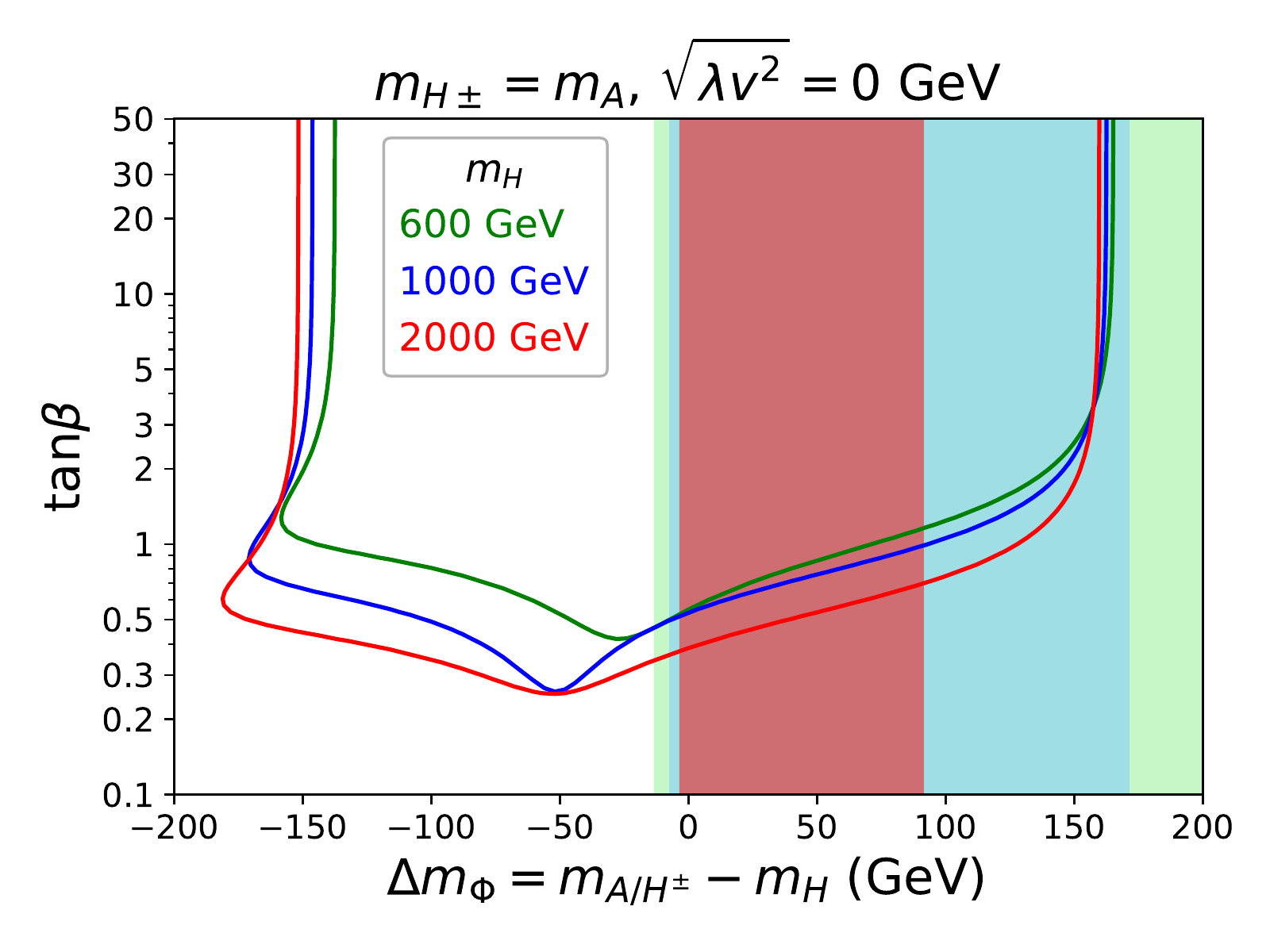}
    \includegraphics[width=0.48\linewidth]{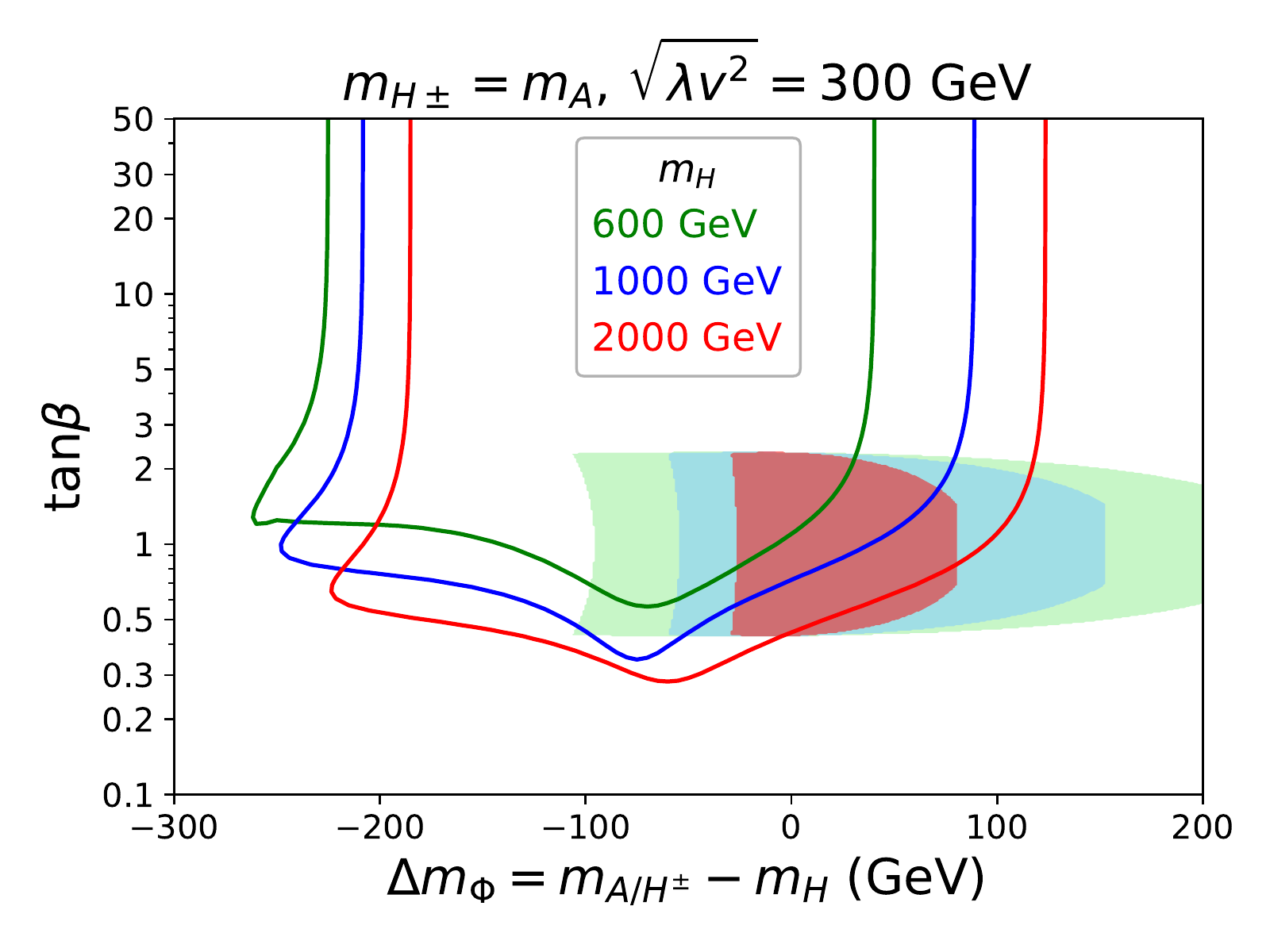}
    \includegraphics[width=0.48\linewidth]{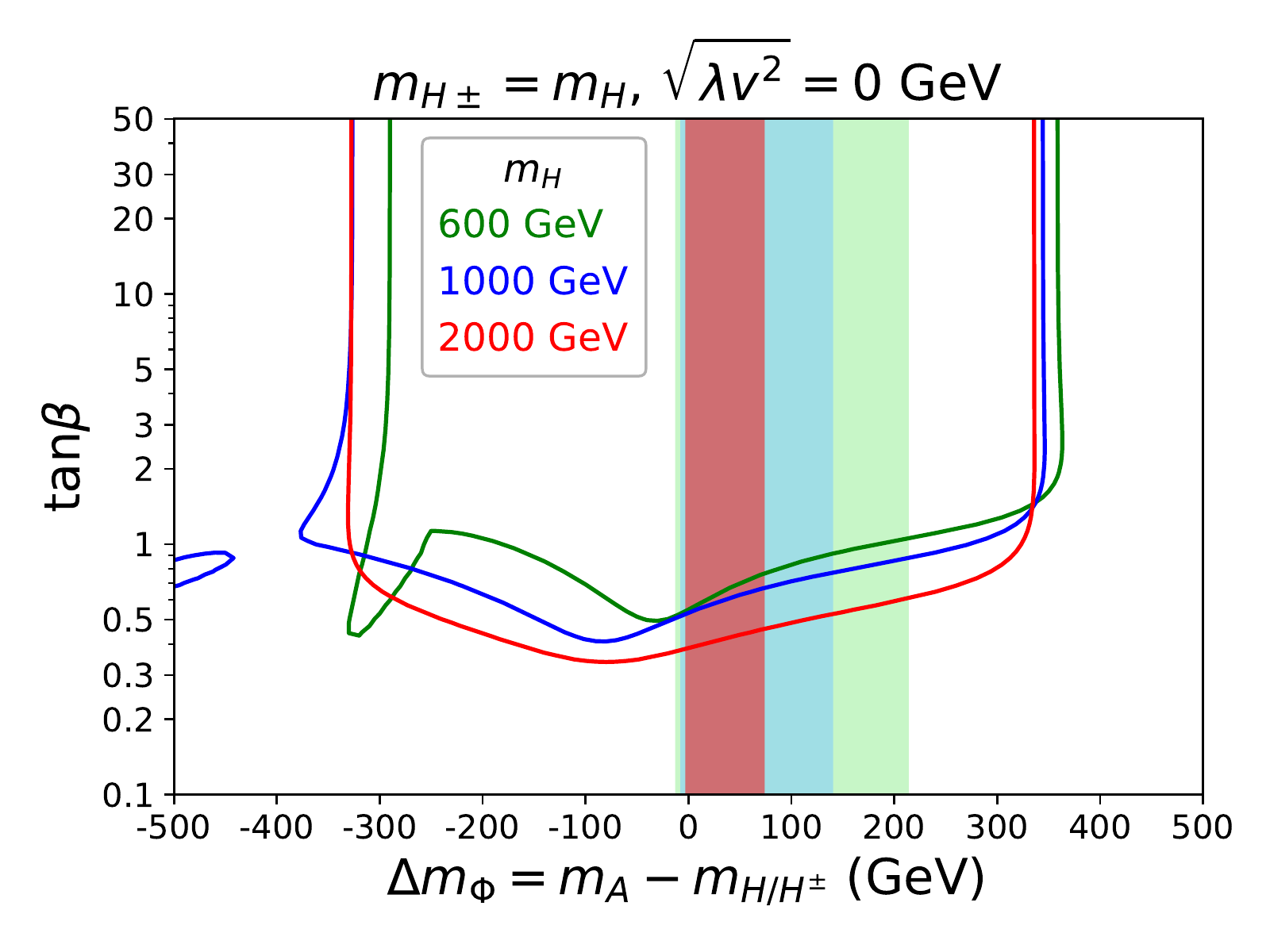}
    \includegraphics[width=0.48\linewidth]{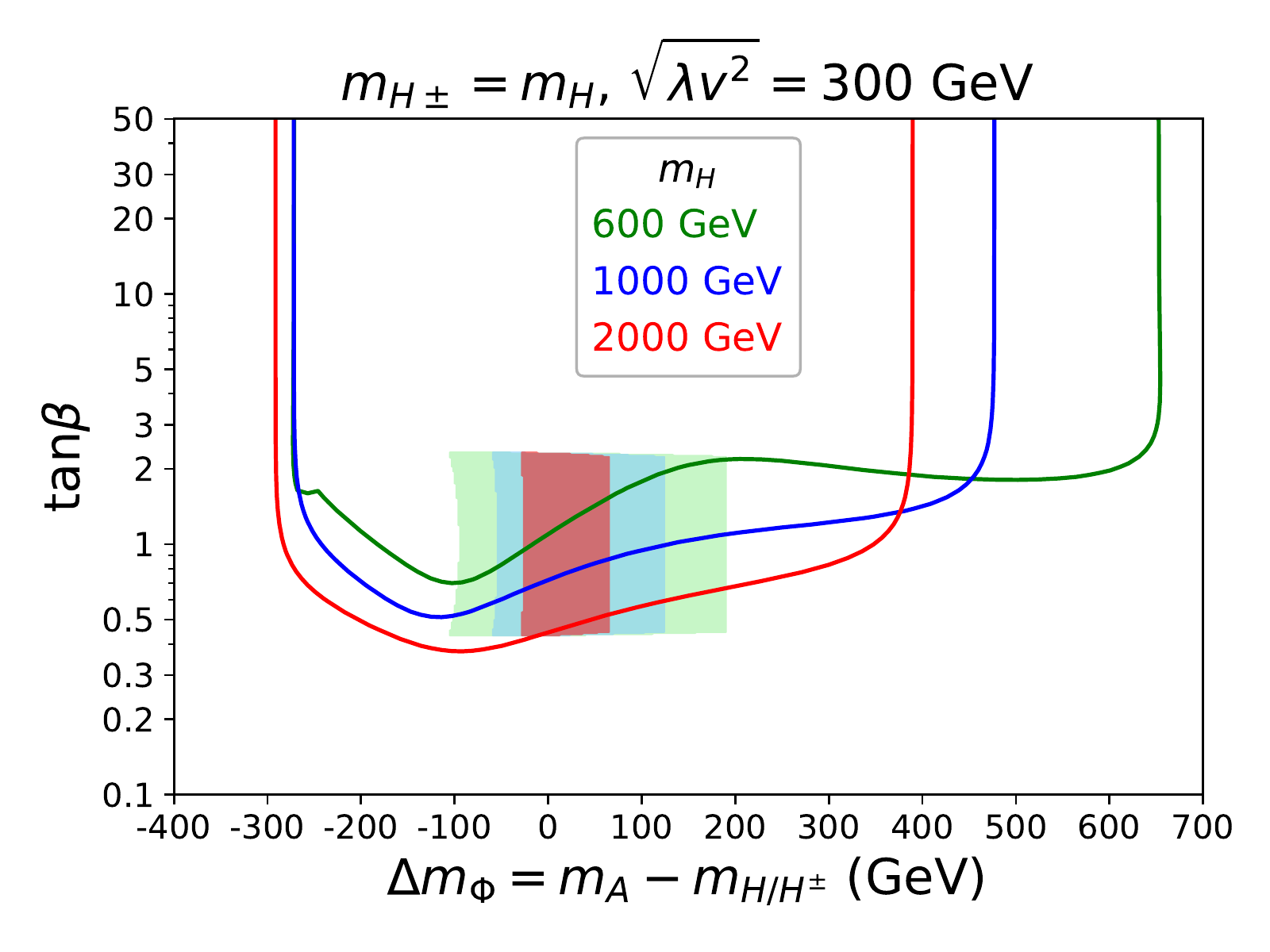}
  \caption{\small Allowed region from  three-parameter fitting results at 95\% C.L. in the 
  $\Delta m_\Phi-\tan\beta$ plane with varying $m_H$ under CEPC precision. Here $\cosba=0$. In the upper two panels, $\Delta m_\Phi=m_{A/H^\pm}-m_H$ and lower two panels, $\Delta m_\Phi=m_A-m_{H/H^\pm}$. $\sqrt{\lambda v^2}$ is set to 0 (left panels) and  300 GeV (right panels). Shaded colored regions are used to indicate theoretical preferred regions.
  }
  \label{fig:tb_dmphi}
\end{figure}

In \autoref{fig:tb_dmphi}, we explore the constrained region in $\Delta  m_\Phi=m_{A/H^\pm}-m_H$ (upper panels) and $\Delta m_\Phi=m_A-m_{H/H^\pm}$ (lower panels) under alignment limit for various values of $m_H$.  $m_{H^\pm} = m_A$ ($m_{H^\pm} = m_H$) is assumed in the former (latter) case to satisfy the $Z$-pole constraints.  
Left and right panels correspond to $\sqrt{\lambda v^2}=0$,  and 300 GeV, respectively.   Shaded colored regions   are used to indicate theoretical preferred regions.

In all panels of \autoref{fig:tb_dmphi}, the allowed regions show a sudden cutoff at certain value of $\Delta m_\Phi$, in particular for $\tan\beta\gtrsim 1$. This feature mainly comes from the loop-level corrections to $\kappa_Z$.  For $m_{H^\pm} = m_A = m_H + \Delta m_\Phi$,  

\begin{align}
\label{eq:kz_cutoff}
\kappa_{Z}^{\rm 1-loop} 
%
%
& \approx 1-\frac{1}{192\pi^2 }\left[\frac{8\lambda_{hHH}^2}{ m_H^2}- 24  \frac{\Delta m_\Phi \lambda_{hHH} }{v^2} + 24\frac{\Delta m_\Phi^2 }{v^2}\right] +\cdots \,.
\end{align}
under the alignment limit and terms proportional to  higher orders of  $\lambda_{hHH}/{m_H}$ are ignored given its typically small size under the alignment limit.  Terms proportional to $\Delta m_\Phi$ are responsible for the sudden cutoff.  For $\lambvs=0$ GeV, $\lambda_{hHH} = -{m_h^2}/{2v}$ is small and $\Delta m_\Phi^2 /{v^2}$ would dominate, resulting in a symmetric bound of $\Delta m_\Phi$.  For a larger value of $\lambvs$, $\Delta m_\Phi \lambda_{hHH} /{v^2}$ becomes more important, resulting in an  asymmetric bound $\Delta m_\Phi$. 
Such behaviour appears for all loop-level Higgs couplings, with $\kappa_Z$ becomes the most constraining one at $\tan\beta \gtrsim 1$.  The case of 
$m_{H^\pm} = m_H = m_A - \Delta m_\Phi$ is similar.
 
\begin{figure}[tb]
\centering
    \includegraphics[width=1\linewidth]{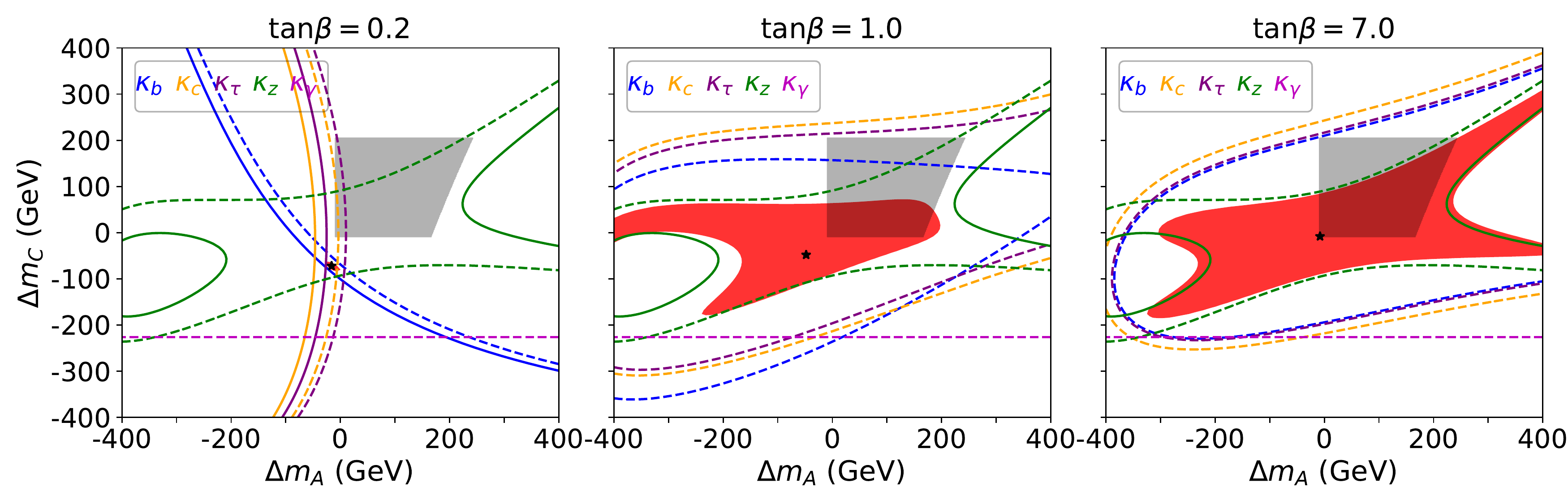}
  \caption{\small Allowed region (red) from two-parameter fitting results at 95\% C.L. in the $\Delta m_A-\Delta m_C$ plane from the individual Higgs coupling measurement, for $\tan\beta=0.2$ (left), 1 (middle), 7 (right) under the alignment limit, with $m_H=800$ GeV and $\sqrt{\lambda v^2}=0$. For individual coupling constraint, the dashed line represents $-\sigma$ limit, while the solid line represents the $+\sigma$ limit. Regions between the solid and dashed curves are the allowed region. For $\kappa_\gamma$, region above the line is allowed. Also shown is the theoretically allowed region shaded in grey color.}
  \label{fig:dmac_ana}
\end{figure}

In \autoref{fig:dmac_ana}, we show the constraints in $\Delta m_C \equiv m_{H^\pm} - m_H$ versus $ \Delta m_A \equiv m_A - m_H$ parameter space, with  
constraints from individual coupling indicated by the colored curves, and the 95\% C.L. $\chi^2$ fit region indicated by the red shaded region with the best fit point indicated by the black star.   Other parameters are chosen as $m_H = 800 \gev$, $\lambvs=0$ and $\tanb=$ 0.2(left), 1(middle), 7(right) under the  alignment limit.   For each individual coupling constraint, the dashed line is for the $-\sigma$ limit, while the solid line is for the $+\sigma$ limit. The range between the two lines is the survival region. Under the alignment limit, $\kappa_Z $ is independent of $\tanb$ as apparent in the figure.  The light grey shadow region is the region preferred by the theoretical considerations.

For the Type-I 2HDM under alignment limit, all the Higgs fermion couplings are $\cot \beta$ enhanced and Higgs-vector boson couplings are $\tanb$-independent. At  small $\tanb$  as shown in the left panel of \autoref{fig:dmac_ana}, $\Delta m_A$ and $\Delta m_C$ are strongly constrained to be close to 0, due to the constraints from $\kappa_b, \kappa_c$ and  $\kappa_\tau$, dominantly.  The blue $\kappa_b$ constraint has a different shape comparing to that of $\kappa_c, \kappa_\tau$, mainly due to the top quark vertex correction.  For larger $\tanb$, the fermion couplings constraints are reduced.  As a result,  $\kappa_Z$ provide the dominant constraint, as shown in the middle and right panels of \autoref{fig:dmac_ana}, which is less constraining.  $\Delta m_{A,C}$ is constrained to be less than about 200 GeV for $\tanb=1$ and less than about 400 GeV for $\tanb=7$, which is quite different from the Type-II 2HDM, which gets tightly constrained in large $\tanb$ as well. For $\tanb >7$, the survived red region does not change significantly.   The typical survived regions for $\Delta m_A, \Delta m_C$ are ($-300$, 400) GeV, ($-200$, 300) GeV respectively at large $\tan\beta$, and generally can be extended to ($-500$,600) GeV and ($-300$, 400) GeV. Compared to the Type-II 2HDM~\cite{Chen:2018shg}, the Type-I 2HDM has a more restricted region at small $\tanb$ and much relaxed region at large $\tanb$.  
%

\begin{figure}[tb]
\centering
    \includegraphics[width=0.48\linewidth]{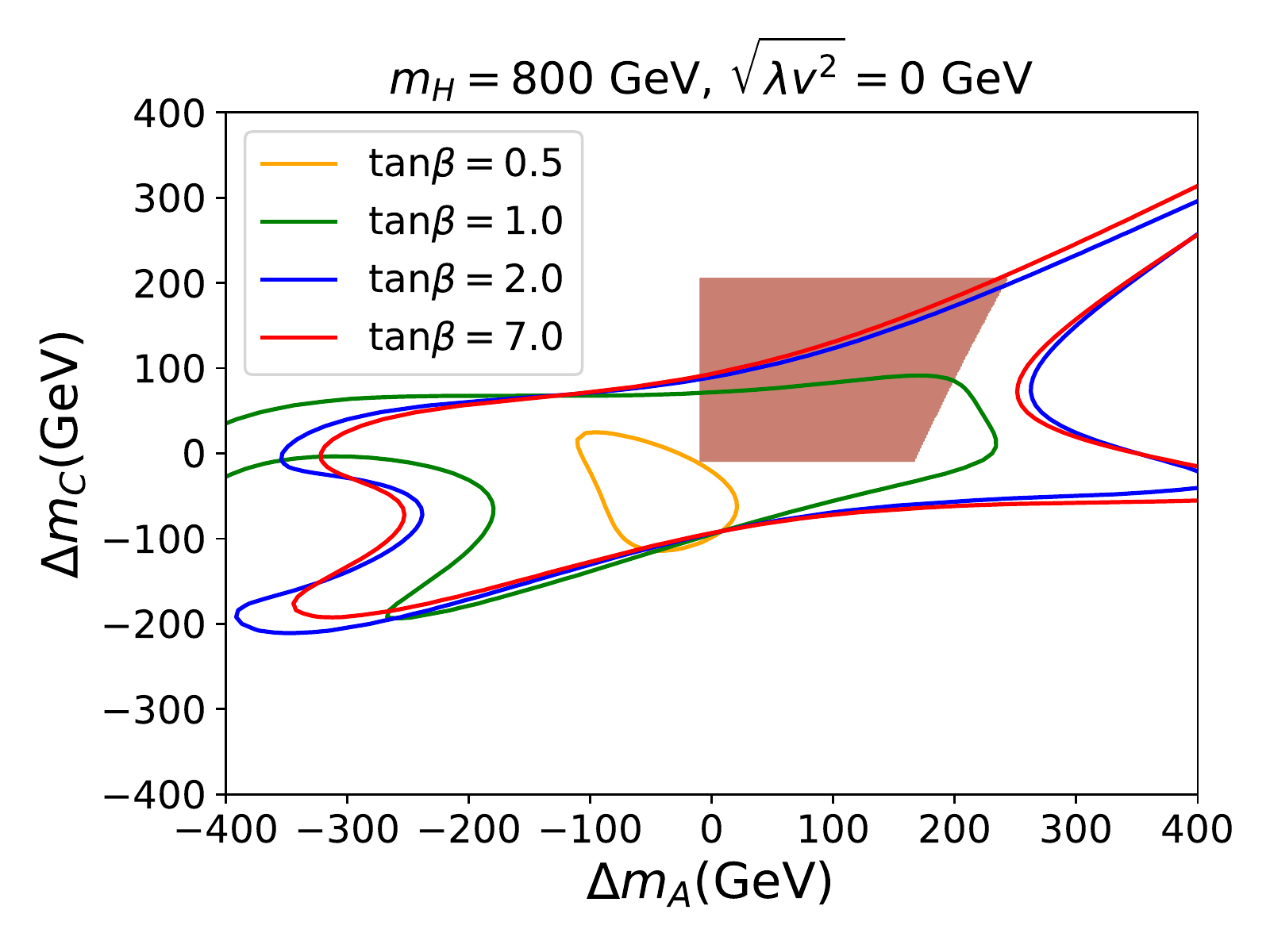}
    \includegraphics[width=0.48\linewidth]{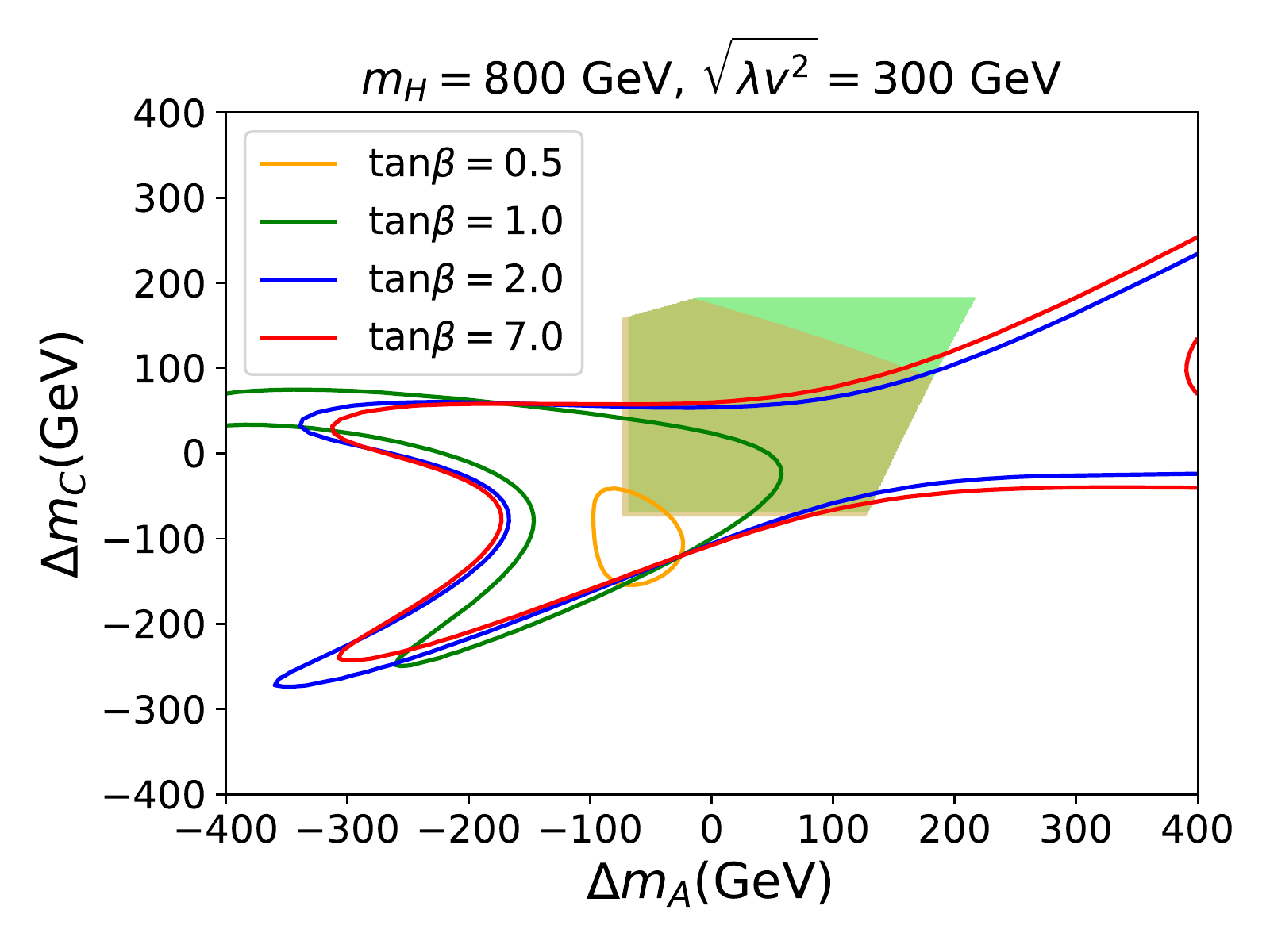}
  \caption{\small Allowed region from  three-parameter fitting results at 95\% C.L. in the $\Delta m_A-\Delta m_C$ plane with varying $\tan\beta$ under $\cosba=0, m_H=800$ GeV. The left panel is for $\sqrt{\lambda v^2}=0$, while the right one for $\sqrt{\lambda v^2}=300$ GeV.  $\tan\beta$ is chosen to be 0.5 (orange), 1 (green), 2 (blue), 7 (red). Also shown are the allowed regions by theoretical constraints, which are indicated with shadows of the same color codes.   }
  \label{fig:dmac_tan}
\end{figure}
In \autoref{fig:dmac_tan}, we show $\chi^2$ fit results at 95\% C.L. in the $\Delta m_A-\Delta m_C$ plane with varying $\tan\beta$ under the alignment limit of $\cosba=0$ for $m_H=800$ GeV.  The left panel is for $\sqrt{\lambda v^2}=0$, and the right panel is  for $\sqrt{\lambda v^2}=300$ GeV.    Also shown in color shaded region are the theoretically preferred regions.   In general, the range for $\Delta m_A$ and $\Delta m_C$ gets bigger for larger $\tan\beta$ and smaller $\lambda v^2$.   In particular, for $\tan\beta=7$, the allowed ranges of $\Delta m_A$ and $\Delta m_C$ shrink to a narrow range around $-100$ GeV.

\begin{figure}[h!]
\centering
    \includegraphics[width=0.48\linewidth]{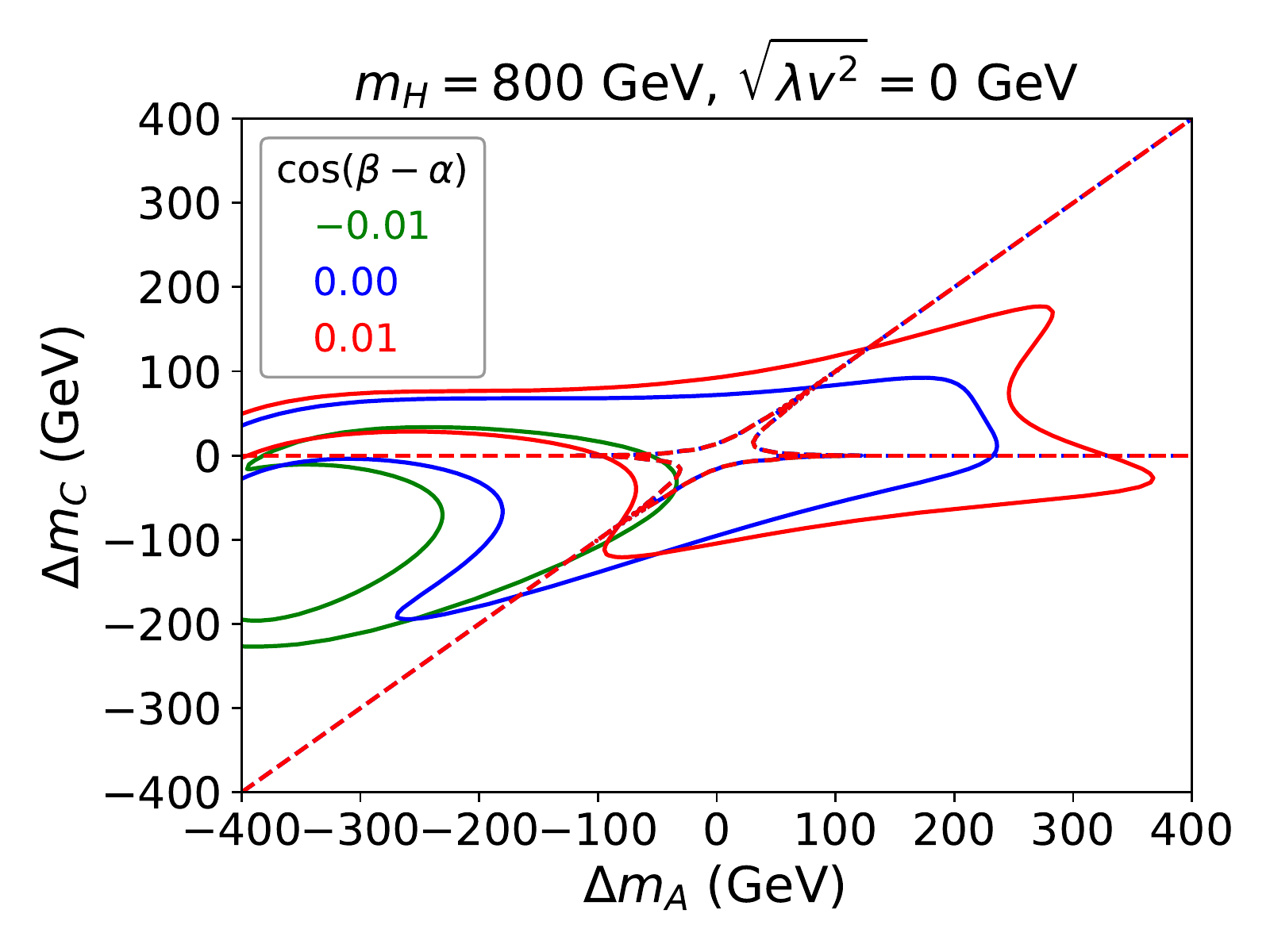}
    \includegraphics[width=0.48\linewidth]{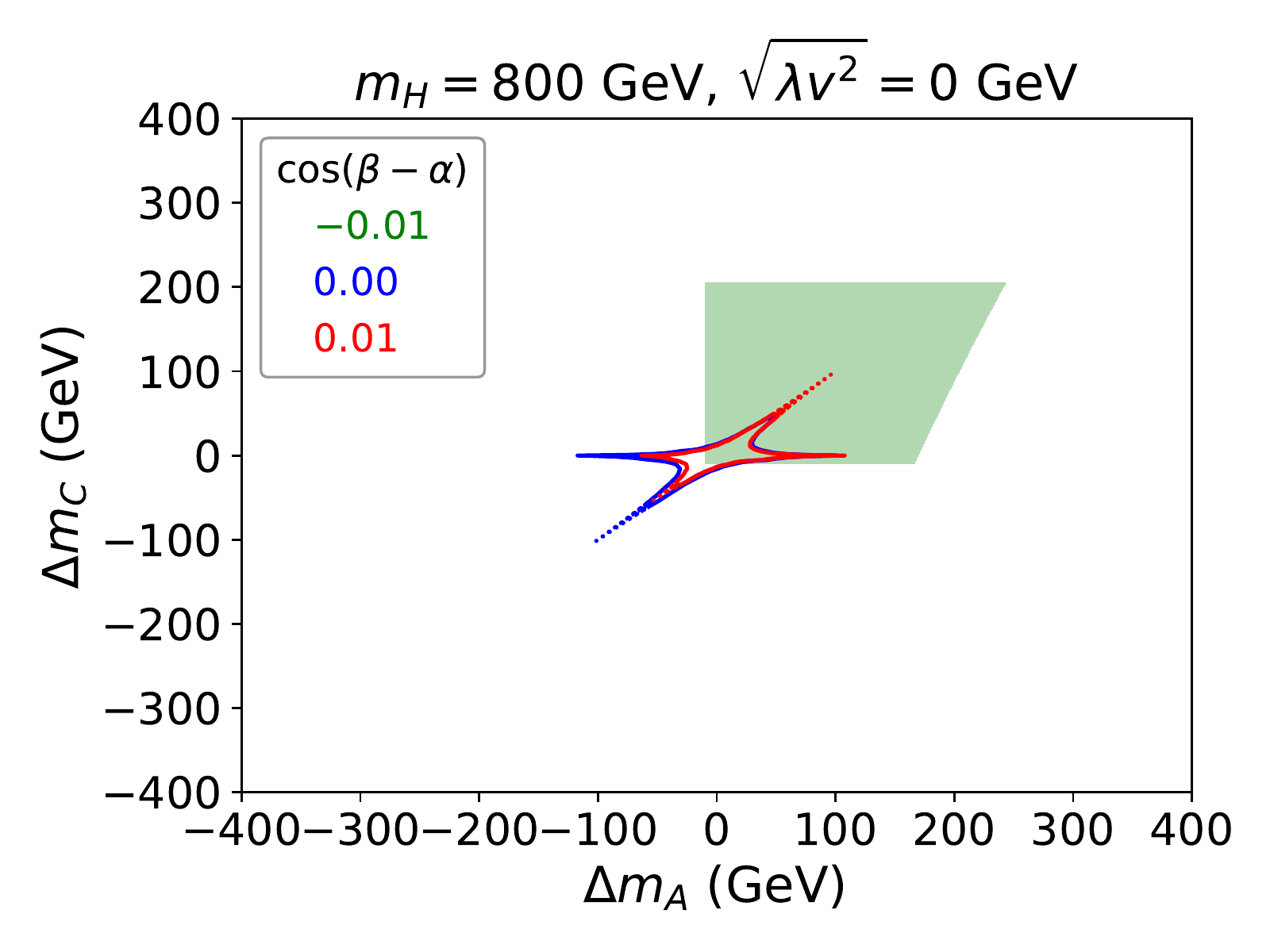}
    \includegraphics[width=0.48\linewidth]{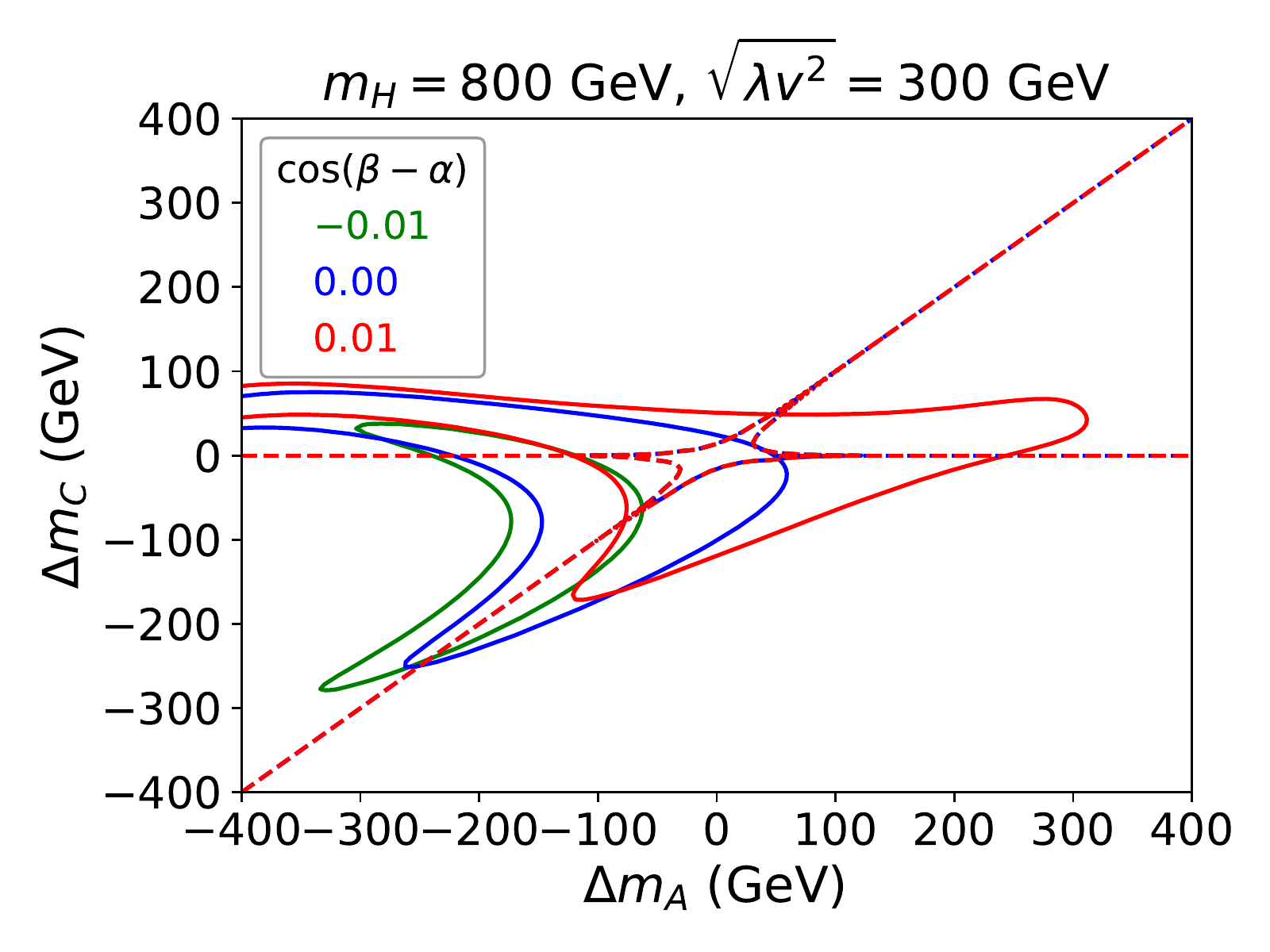}
    \includegraphics[width=0.48\linewidth]{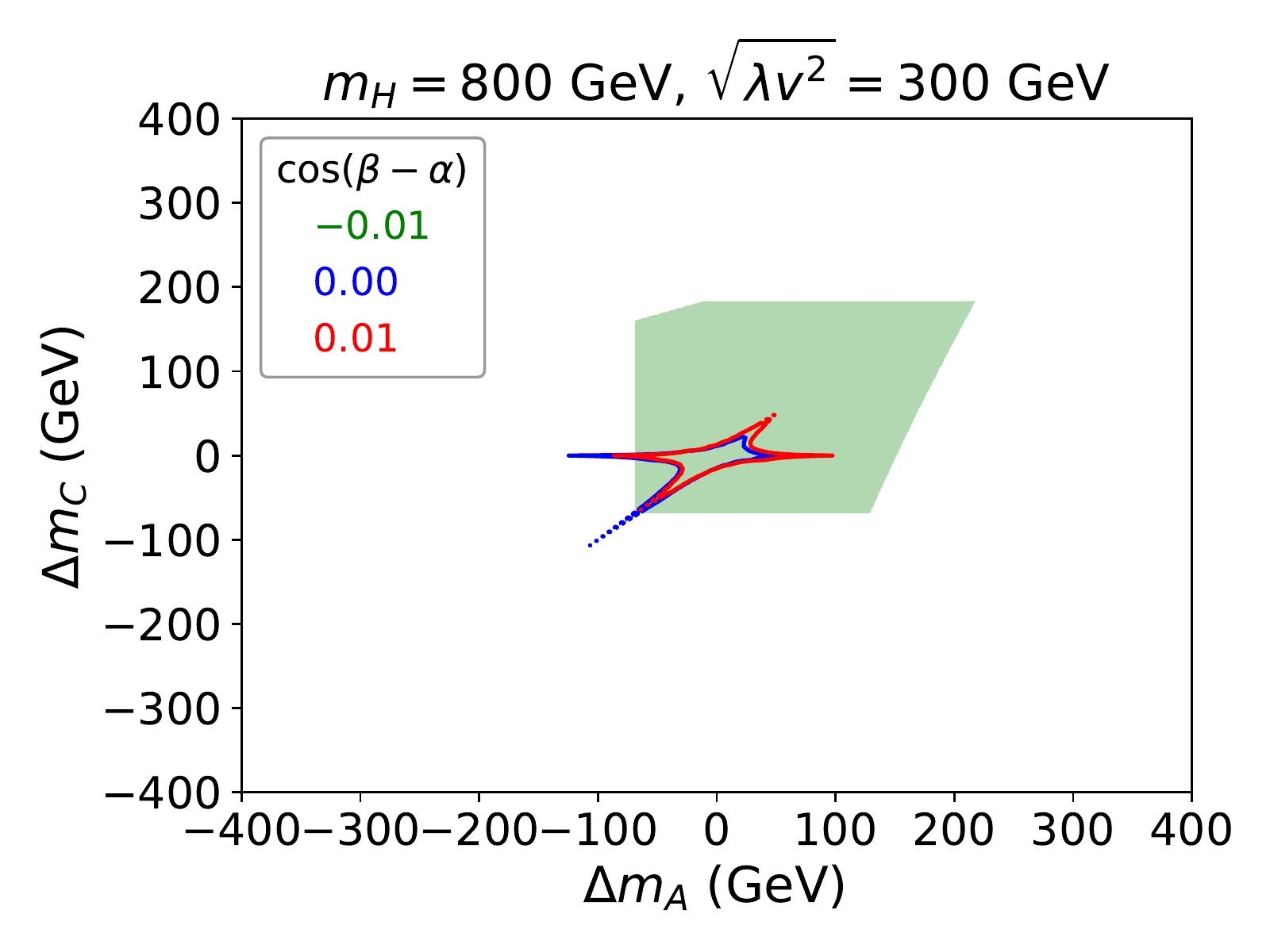}
    \includegraphics[width=0.48\linewidth]{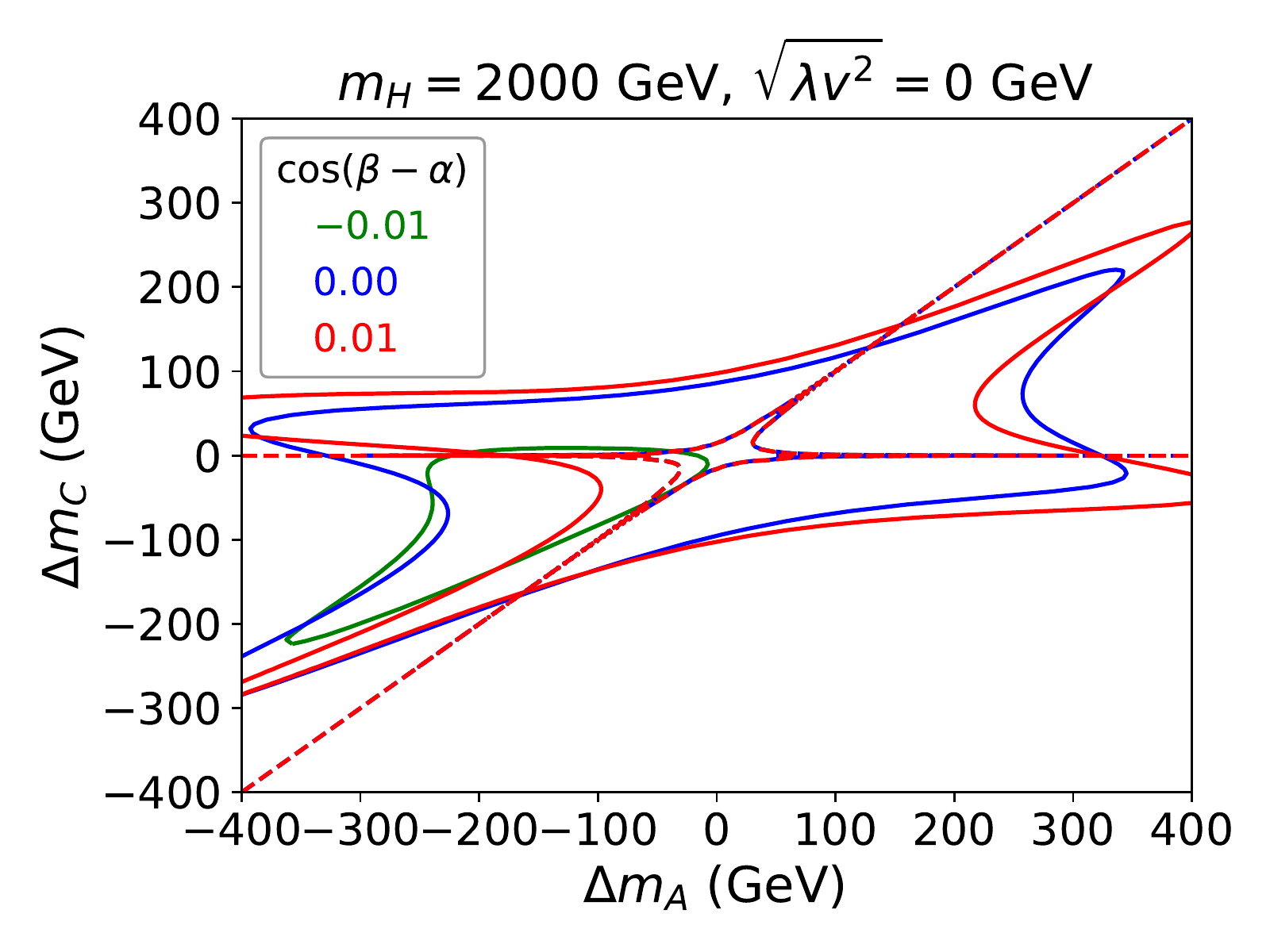}
    \includegraphics[width=0.48\linewidth]{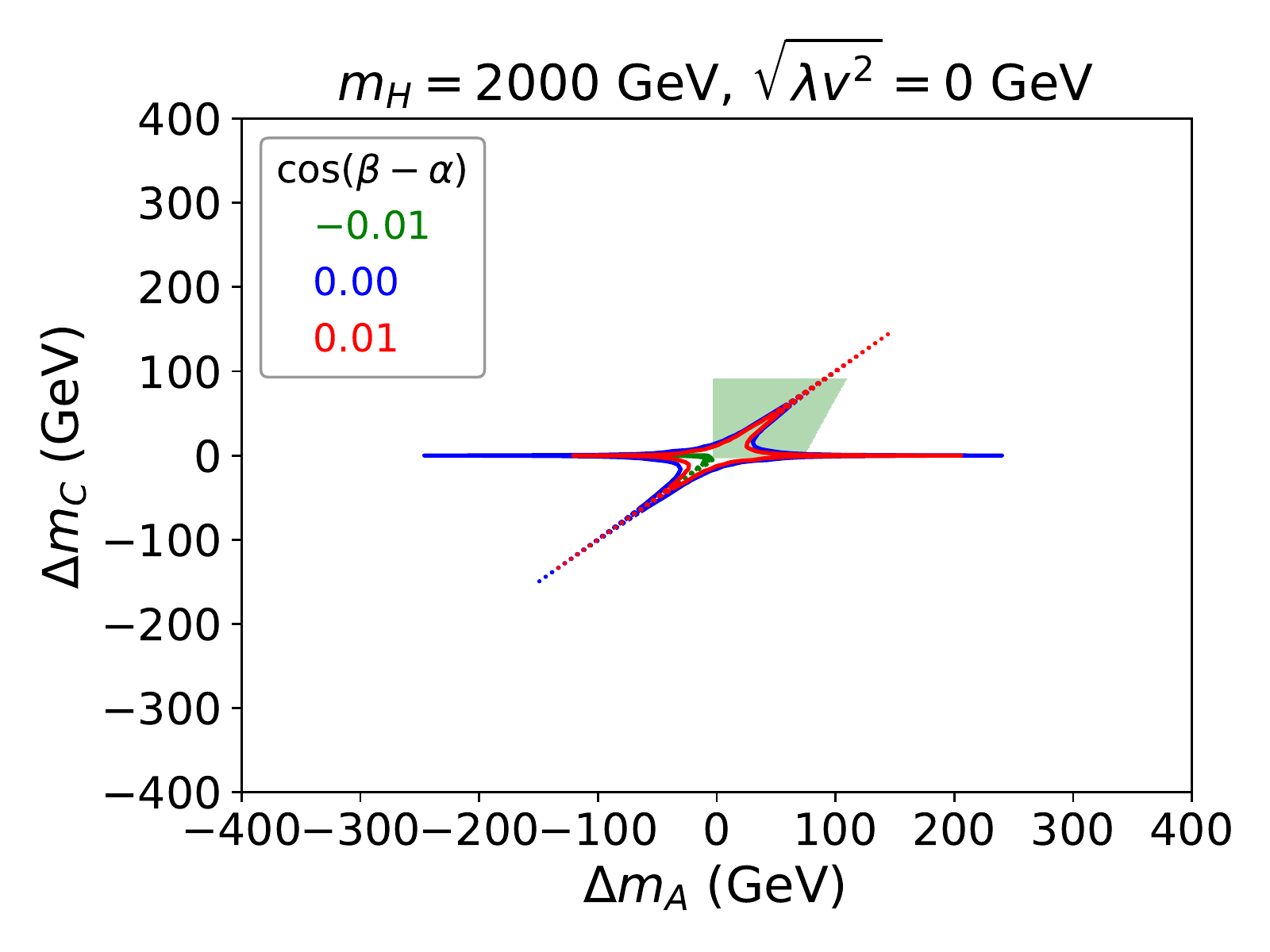}
 \caption{\small Allowed region from  three-parameter fitting results at 95\% C.L. in the $\Delta m_A-\Delta m_C$ plane and also varying  $\cos(\beta-\alpha)$, for Higgs (solid curve) and $Z$-pole (dashed curve) constraints (left panel), and combined constraints (right panels), with upper row for $m_H=600$ GeV, $\sqrt{\lambda v^2}=0$, middle row for $m_H=600$ GeV, $\sqrt{\lambda v^2}=300$ GeV, and bottom row for $m_H=2000$ GeV, $\sqrt{\lambda v^2}=0$. $\tan\beta=1$ is assumed for all plots. Shaded region is used to indicate the theoretical constraints at $\cos(\beta-\alpha)=0$. }
  \label{fig:da-dc-cos}
\end{figure}


In \autoref{fig:da-dc-cos}, we take $Z$-pole precision into account as well.  The left panels show Higgs precision (solid curves) and $Z$-pole precision (dashed curves) for different $\cos(\beta-\alpha)$ values: $-0.01$ (green curve), 0 (blue curve) and 0.01 (red curve). The right panels show the combined fitting results.  Shaded region is used to indicate the theoretical constraints at $\cos(\beta-\alpha)=0$.  While the $Z$-pole precision measurements are more constraining in the mass difference of $\Delta m_A$ and $\Delta m_C$ in general, they can always be satisfied for $\Delta m_C=0$ ($m_H^\pm =m_H$) or $\Delta m_A=\Delta m_C$ ($m_H^\pm = m_A$).  The Higgs precision measurements, on the other hand, could provide an upper limit on $|\Delta m_{A,C}|$.  When combined together, a more restrictive range of $\Delta m_{A,C}$ can be achieved.

\subsection{Comparison between different lepton colliders}

\begin{figure}[h!]
\centering
    \includegraphics[width=0.48\linewidth]{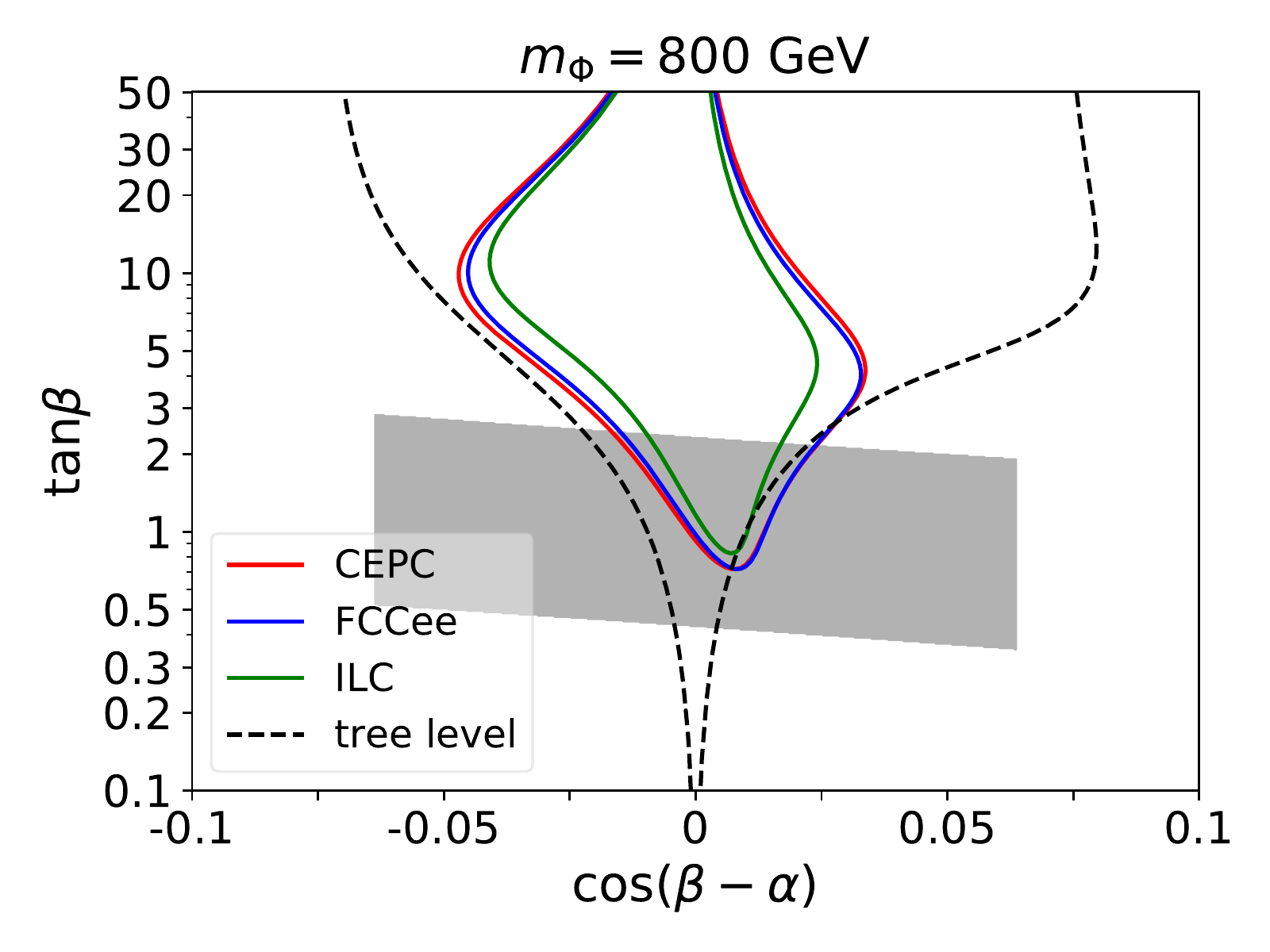}
    \includegraphics[width=0.48\linewidth]{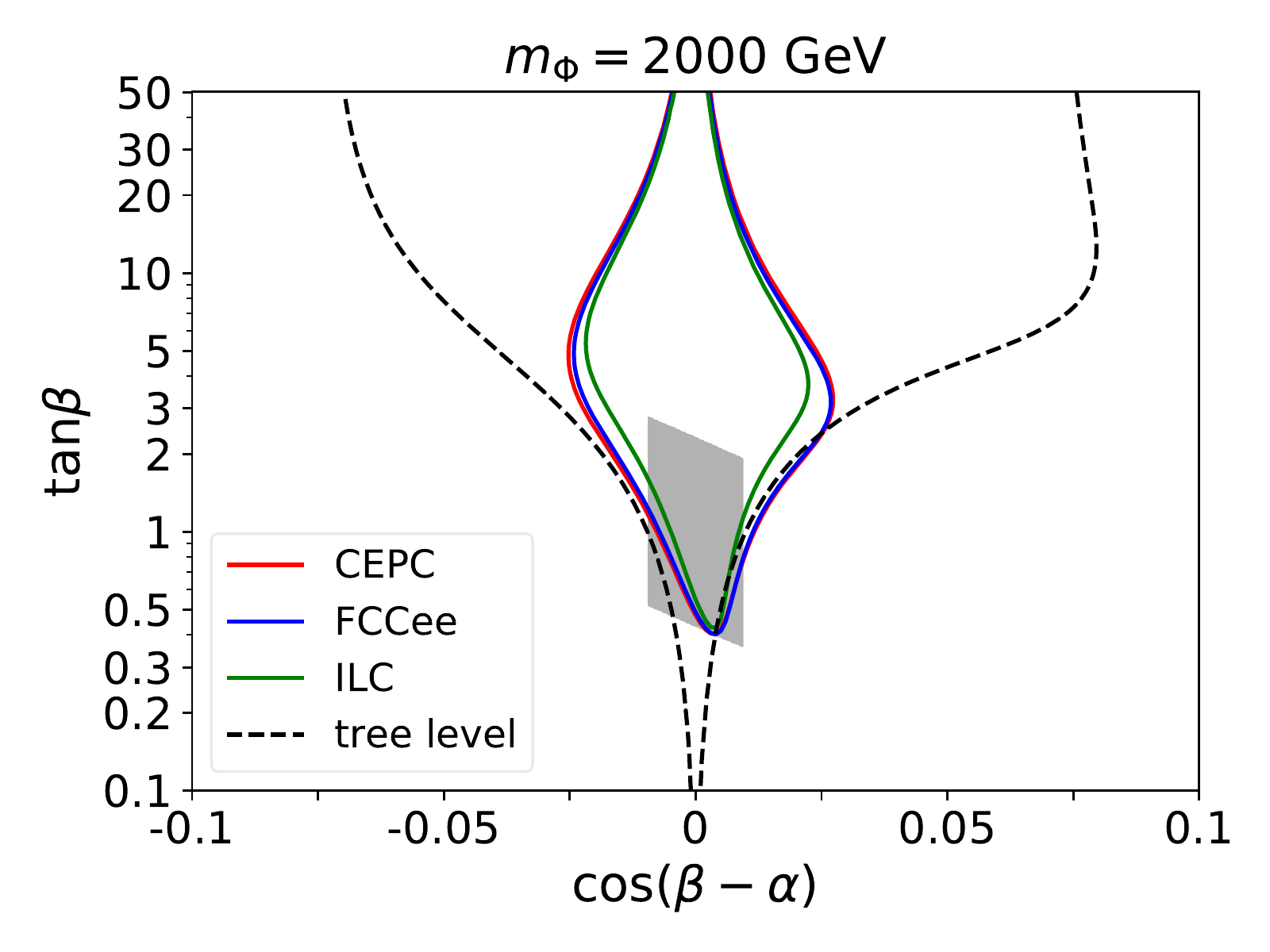}
  \caption{\small Allowed region from  two-parameter fitting results at 95\% C.L. in the $\cos(\beta-\alpha)-\tan\beta$ plane with CEPC (red), FCC-ee (blue) and ILC (green) precision. The black dashed line indicates the CEPC tree-level only results as a comparison. $\sqrt{\lambda v^2}$ is set to be 300 GeV,  with $m_\Phi$ = 800 GeV (left panel),  and 2000 GeV (right panel). The grey shadow indicates the survival region of theoretical constraint.}
  \label{fig:ee-costanb}
\end{figure}

To compare the sensitivities of different Higgs factory machine options, in Fig.~\ref{fig:ee-costanb}, we show the reach in $\cos(\beta-\alpha)- \tan\beta$ plane for CEPC (red curve), FCC-ee (blue curve) and ILC (green curve) for $m_\phi=800$ GeV (left panel) and 2000 GeV (right panel). 
Tree level results with CEPC precision are indicated in black dashed line to guide the eye.   
The reach of CEPC and FCC-ee is similar, while ILC has slightly better reach given the various center of mass energy options.

\begin{figure}[h!]
\centering
    \includegraphics[width=0.48\linewidth]{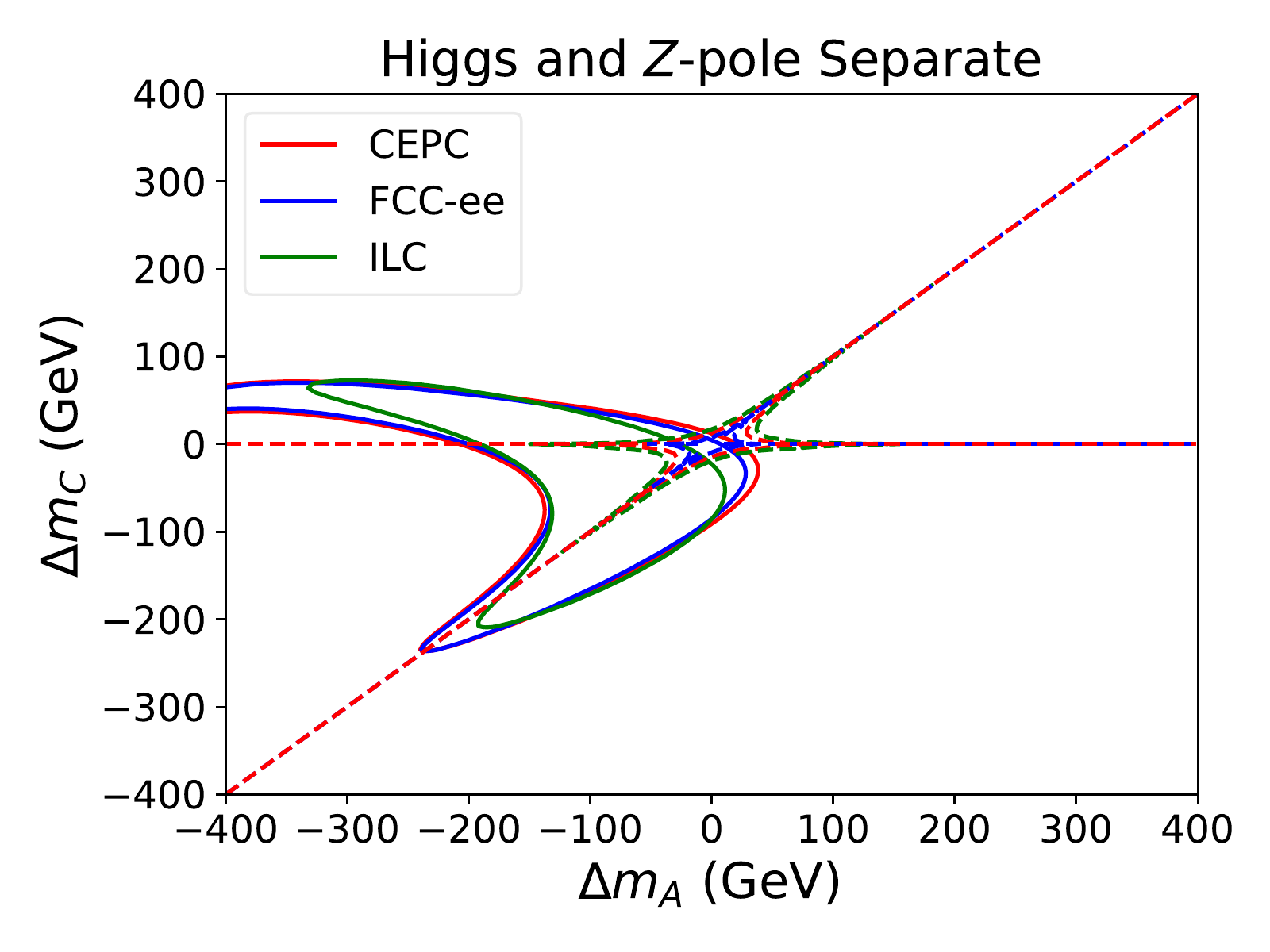}
    \includegraphics[width=0.48\linewidth]{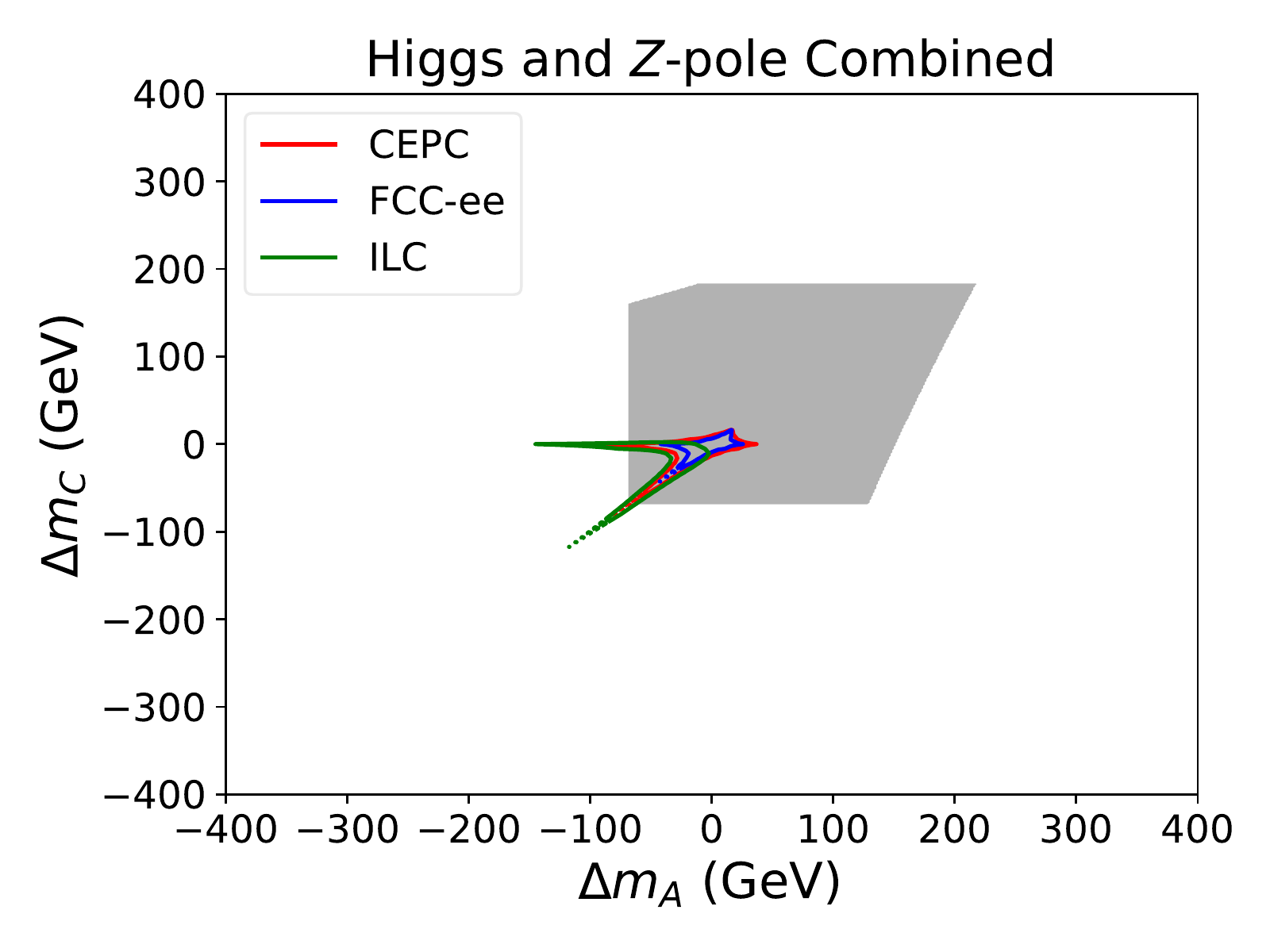}
\caption{\small Allowed region from  two-parameter fitting results at 95\% C.L. in the $\Delta m_A-\Delta m_C$ plane with CEPC (red), FCC-ee (blue) and ILC (green) precisions.  The left and right panels are for Higgs/$Z$-pole results individually and combined, respectively. Here $m_H$ = 800 GeV, $\sqrt{\lambda v^2}$ = 300 GeV, $\cos(\beta-\alpha)=0$, $\tan\beta=1$. In the left panel, Higgs precision measurement result is given by solid line while $Z$-pole precison measurement result is given by dashed line. The grey shadow in the right panel indicates the survival region of theoretical constraint.  }
  \label{fig:ee-dmac}
\end{figure}

In Fig.~\ref{fig:ee-dmac}, we show the 95\% C.L. constraints on the $\Delta m_A-\Delta m_C$ plane with both Higgs and $Z$-pole precision measurements under alignment limit. 
Left panel is for the individual constraints while the right panel show the combined fit. 
While ILC has better Higgs precision reach, FCC-ee is slightly better for $Z$-pole reach. 
Combined together, reaches of three machine options are similar, and the typical allowed mass splitting  is about 200 GeV.


\section{Summary and Conclusions}
\label{sec:conclu}

With the discovery of the SM-like Higgs boson at the LHC, searching for additional Higgs bosons beyond the SM is strongly motivated from both theoretical and experimental considerations. 
In the absence of signals from the direct searches in the LHC experiments, it would be prudent to seek for complementary means,  in particular,  from the precision measurements of the SM observables which are sensitive to BSM new physics.
In this paper, extending the previous work on the Type-II 2HDM \cite{Chen:2018shg}, we performed a comprehensive study for the Type-I 2HDM from the impacts of the precision measurements of the SM observables at the proposed $Z$-factories and Higgs factories on the extended Higgs sector.

First, we listed the latest expected accuracies on determining the EW observables at the $Z$-pole and the Higgs factories (Table~\ref{tab:mu_precision} in  \autoref{sec:input}), as a general guidance and inputs for the following studies. We gave a brief summary for Type-I 2HDM in~\autoref{sec:model} to specify the model and set the scope for the rest of the paper by introducing the degenerate and non-degenerate cases. We then presented the existing constraints on the model parameters from theoretical considerations~(\autoref{fig:theory_case2} in~\autoref{sec:theory}) and the LHC bounds from the current searches and the future expectations~(\autoref{fig:TypeIAHExcl} and~\autoref{fig:TypeIAhZ} in~\autoref{sec:LHC}).
Previous works focused on either just the tree-level deviations, or loop corrections under the alignment limit, and with the assumption of degenerate masses of the heavy Higgs bosons. A recent study~\cite{Chen:2018shg} extended the previous work to have included the general one-loop effects in the Type-II 2HDM.
In our analyses, we extended the existing results by including the tree-level and one-loop level effects of non-degenerate Higgs masses in the Type-I 2HDM.

The main results of the paper were presented in~\autoref{sec:results}, where we performed a $\chi^2$ fit to the expected precision measurements in the full model-parameter space.  We first illustrated the simple case with degenerate heavy Higgs masses as in~\autoref{fig:tanbcba_fit_degenerate} with the expected CEPC precision.  We found that  in the parameter space of $\cos(\beta-\alpha)$ and $\tan\beta$, the largest 95\% C.L. range of $|\cos(\beta-\alpha)|\lesssim 0.05$ could be achieved for $\tan\beta $ between $5-10$, with smaller and larger values of $\tan\beta$ tightly constrained by $\kappa_{g,c,b,\tau}$ and $\kappa_{Z}$, respectively. For the Type-I 2HDM, large $\tanb$ regions were always less restricted since all Yukawa couplings are $\cot \beta$-enhances. When including loop-level corrections, the large $\tanb$ regions got additional $\tanb$-enhanced constraints from triple Higgs couplings terms proportional to $\cosba \tanb$ as in \autoref{eq:loop_tan}.
Varying heavy Higgs masses and   $\lambda v^2$, as shown in~\autoref{fig:tanb_cba_sum}, the significant loop-level effect shown again and shifted the 95\% C.L. region. The low $\tan\beta$ results in Figs.~\ref{fig:tanbcba_fit_degenerate} and \ref{fig:tanb_cba_sum} are similar 
to those in Type-II, but constraints are stronger since all couplings are $\cot \beta$-enhanced around $\cosba=0$.

The limits on the heavy Higgs masses also depend on $\tan\beta$, $\lambda v^2$ and $\cos(\beta-\alpha)$, as shown in~\autoref{fig:tbvsmass} and alternatively in~\autoref{fig:tbvsmassm12} varying $m_{12}^2$.  While the most relaxed limits can be obtained under $\cosba=0$ with small $\lambda v^2$, deviation away from $\cosba=0$ leads to tighter constraints, especially for the  allowed range of $\tan\beta$.  The reach seen in the $m_\Phi-\tan\beta$ plane is complementary to direct non-SM Higgs search limits at the LHC and future $pp$ colliders as in \autoref{fig:TypeIAHExcl}, especially in the large $\tan\beta$ region when the direct search limits are relaxed.

It is important to explore the extent to which the parametric deviations from the degenerate mass case can be probed by the precision measurements.
Figure \ref{fig:tb_dmphi} showed the allowed deviation for $\Delta m_\Phi$ with the expected CEPC precision and~\autoref{fig:dmac_ana} demonstrated the constraints from the individual decay channels of the SM Higgs boson. As shown in~\autoref{fig:dmac_tan}, the Higgs precision measurements alone constrain $\Delta m_{A,C}$ to be less than about a few hundred GeV, with tighter constraints achieved for
small $m_H$, large $\lambda v^2$ and small values of $\tan\beta$.   $Z$-pole measurements, on the other hand, constrain the deviation from $m_{H^\pm}\sim m_{A,H}$.  We found that the expected accuracies at the $Z$-pole and at a Higgs factory are quite complementary in constraining mass splittings.
While $Z$-pole precision is more sensitive to the mass splittings between the charged Higgs and the neutral ones (either $m_H$ or $m_A$),  Higgs precision measurements in addition could impose an upper bound on the mass splitting between the neutral ones. Combining both Higgs and $Z$-pole precision measurements,  the mass splittings are constrained even further, as shown in~\autoref{fig:da-dc-cos}, especially when deviating from the alignment limit.   In summary, Higgs precision measurements are more sensitive to parameters like $\cos(\beta-\alpha)$, $\tan\beta$, $\sqrt{\lambda v^2}$ and the masses of heavy Higgs bosons.   We found that except for cancellations in some correlated parameter regions, the allowed ranges are typically
 \begin{equation}
\tan\beta \geq 0.3,\quad |\cos(\beta-\alpha)| < 0.05, \quad  |\Delta m_\Phi | < 200\ {\rm GeV}\,.
 \end{equation}

We mostly presented our results adopting the expectations of the CEPC precision on Higgs and $Z$-pole measurements.  The comparison among different proposed Higgs factories of CEPC, FCC-ee and ILC are illustrative and are shown in Figs.~\ref{fig:ee-costanb} and \ref{fig:ee-dmac}.   While the ILC with different center-of-mass energies has slightly better reach in the Higgs precision fit, the FCC-ee has slightly better reach in the $Z$-pole precision fit.

While the precision for the Higgs coupling measurements with the LHC program is expected to be at the order of few percent, the precision measurements of the SM observables at the proposed $Z$ and Higgs factories would significantly advance our understanding of the electroweak physics and shed lights on possible new physics beyond the SM, and could therefore be complementary to the direct searches at the LHC and future hadron colliders.


\begin{acknowledgments}


We would also like to thank Liantao Wang and Manqi Ruan for valuable discussions.
NC is supported by the National Natural Science Foundation of China (under Grant No. 11575176) and Center for Future High Energy Physics (CFHEP).
TH is supported in part by the U.S.~Department of Energy under grant No.~DE-FG02-95ER40896 and by the PITT PACC. SS is supported  by the Department of Energy under Grant No.~DE-FG02-13ER41976/DE-SC0009913.
WS were supported by  the  Australian  Research  Council  Discovery  Project DP180102209. YW is supported by the Natural Sciences and Engineering Research Council of Canada (NSERC). TH also acknowledges the hospitality of the Aspen Center for Physics, which is supported by National Science Foundation grant PHY-1607611.

 \end{acknowledgments}

\appendix
\section{Analytic Calculation}
\label{sec:app}

While the analysis is based on the full expressions of each loop corrections, some analytical formulae (with possible approximation) can be useful to provide some physical insights. The most relevant couplings are $\kappa_b$ and $\kappa_Z$. In this appendix, we list the dominant contributions in the loop correction results for these two cases. Note that, the expressions are only valid in Type-I case.  

 The most important contributions to $\kappa_b$ come from two parts: (1) the top yukawa coupling (which connects with $b$ quark by SU(2)$_L$ symmetry), (2) the triple Higgs couplings.
By only keeping parts relevant to above two contributions, we have
\begin{align}
\kappa_b=&\xi_h^b+\frac{1}{16\pi^2}\biggl\{\frac{2m_t^2\xi_h^t\cot^2\beta}{v^2}\biggl[m_h^2C_{12}(m_b^2,m_b^2,m_h^2;m_{t}^2,m_{H^\pm}^2,m_{t}^2)\nonumber \\
&+2m_{t}^2C_0(m_b^2,m_b^2,m_h^2;m_{t}^2,m_{H^\pm}^2,m_{t}^2)+v\lambda_{hH^+H^-}C_0(m_b^2,m_b^2,m_h^2;m_{H^\pm}^2,m_{t}^2,m_{H^\pm}^2)\biggr]\nonumber\\
&-\frac{\xi_h^b}{2}\biggl[\lambda_{H^+H^-h}^2\frac{d}{dp^2}B_0(m_h^2,m_{H^{\pm}}^2,m_{H^{\pm}}^2)+2\lambda_{AAh}^2\frac{d}{dp^2}B_0(m_h^2,m_A^2,m_A^2)\nonumber\\
&+2\lambda_{HHh}^2\frac{d}{dp^2}B_0(m_h^2,m_H^2,m_H^2)\biggr]+\frac{3m_t^2\xi_h^b\cot^2\beta}{v^2}B_0(m_A^2,m_t^2,m_t^2)\nonumber\\
&-\frac{3m_t^4\xi_h^t\xi_H^t\xi_H^b}{(m_H^2-m_h^2)v^2}\biggl[\biggl(4-\frac{m_h^2}{m_t^2}\biggr)B_0(m_h^2,m_t^2,m_t^2)-\biggl(4-\frac{m_H^2}{m_t^2}\biggr)B_0(m_H^2,m_t^2,m_t^2)\biggr]\nonumber\\
&+\frac{\xi_H^b}{2(m_H^2-m_h^2)}\biggl[\lambda_{H^+H^-h}\lambda_{H^+H^-H}\biggl(B_0(m_h^2,m_{H^\pm}^2,m_{H^\pm}^2)-B_0(m_H^2,m_{H^\pm}^2,m_{H^\pm}^2)\biggr)\nonumber\\
&+2\lambda_{AAh}\lambda_{AAH}\biggl(B_0(m_h^2,m_{A}^2,m_{A}^2)-B_0(m_H^2,m_{A}^2,m_{A}^2)\biggr)\nonumber\\
&+6\lambda_{HHh}\lambda_{HHH}\biggl(B_0(m_h^2,m_{H}^2,m_{H}^2)-B_0(m_H^2,m_{H}^2,m_{H}^2)\biggr)
\biggr]
\biggr\}
\end{align}
where $B_0$, $C_0$ and $C_{12}$ are Passarino-Veltman functions in {\tt LoopTools}~\cite{Hahn:1998yk} convention.
 
The main contributions in $\kappa_Z$ only come from the triple Higgs couplings:
\begin{align}
  \kappa_{Z}=&\xi_h^Z-\frac{1}{32\pi^2}\biggl[\lambda_{hH^+H^-}^2\frac{d}{dp^2}B_0(m_h^2,m_{H^\pm}^2,m_{H^\pm}^2)\nonumber\\
  &+2\lambda_{hAA}^2\frac{d}{dp^2}B_0(m_h^2,m_A^2,m_A^2)+2\lambda_{hHH}^2\frac{d}{dp^2}B_0(m_h^2,m_{H}^2,m_{H}^2)\biggr]\nonumber\\
  &-\frac{1}{32\pi^2 v}\biggl[2\cos^22\theta_W\lambda_{hH^+H^-}B_0(m_h^2,m_{H^\pm}^2,m_{H^\pm}^2)+2\lambda_{hAA}B_0(m_h^2,m_A^2,m_A^2)\nonumber\\
  &+2\lambda_{hHH}B_0(m_h^2,m_H^2,m_H^2) -8\cos^22\theta_W\lambda_{hH^+H^-}C_{24}(m_Z^2,m_Z^2,m_h^2,m_{H^\pm}^2,m_{H^\pm}^2,m_{H^\pm}^2)\nonumber\\
  &-8\lambda_{hAA}C_{24}(m_Z^2,m_Z^2,m_h^2,m_{A}^2,m_{H}^2,m_{A}^2)-8\lambda_{hHH}C_{24}(m_Z^2,m_Z^2,m_h^2,m_{H}^2,m_{A}^2,m_{H}^2)\biggr]
\end{align}

$\lambda$'s in both $\kappa_b$ and $\kappa_Z$ are the triple Higgs couplings:
\begin{align}
\lambda_{H^+H^-h}=&-\frac{1}{v}\biggl[(2M^2-2m_{H^{\pm}}^2-m_h^2)s_{\beta-\alpha}+2(M^2-m_h^2)\cot2\beta c_{\beta-\alpha}\biggr]\\
\lambda_{AAh}=&-\frac{1}{2v}\biggl[(2M^2-2m_{A}^2-m_h^2)s_{\beta-\alpha}+2(M^2-m_h^2)\cot2\beta c_{\beta-\alpha}\biggr]\\
\lambda_{HHh}=&-\frac{s_{\beta-\alpha}}{2v}\biggl[(2M^2-2m_{H}^2-m_h^2)s^2_{\beta-\alpha}\nonumber\\
&+2(3M^2-2m_H^2-m_h^2)\cot2\beta s_{\beta-\alpha}c_{\beta-\alpha}-(4M^2-2m_H^2-m_h^2)c_{\beta-\alpha}^2\biggr]\\
\lambda_{H^+H^-H}=&\frac{1}{v}\biggl[2(M^2-m_H^2)\cot2\beta s_{\beta-\alpha}+(2m_{H^\pm}^2+m_H^2-2M^2)c_{\beta-\alpha}\biggr]\\
\lambda_{AAH}=&\frac{1}{2v}\biggl[2(M^2-m_H^2)\cot2\beta s_{\beta-\alpha}+(2m_{A}^2+m_H^2-2M^2)c_{\beta-\alpha}\biggr]\\
\lambda_{HHH}=&\frac{1}{2v}\biggl[2(M^2-m_H^2)\cot2\beta s_{\beta-\alpha}^3-2(M^2-m_H^2)c_{\alpha-\beta}s_{\beta-\alpha}^2+m_H^2c_{\beta-\alpha}\biggr]
\end{align}
where $M^2=\frac{m_{12}}{\cos\beta\sin\beta}$.

\bibliographystyle{JHEP}
\bibliography{references}

\end{document}